\documentclass[preprint,12pt]{elsarticle}

\usepackage{lineno,hyperref}
\usepackage{amsmath}
\usepackage{graphicx}
\usepackage{subcaption}

\journal{NIM A}

\begin{document}

\begin{frontmatter}

\title{CsI(Tl) Pulse Shape Discrimination with the Belle II Electromagnetic Calorimeter as a Novel Method to Improve Particle Identification at Electron-Positron Colliders}

\address[a:uvic] {University of Victoria, Victoria, British Columbia, V8W 3P6, Canada}
\address[a:DESY] {Deutsches Elektronen--Synchrotron, 22607 Hamburg, Germany}
\address[a:NaraWu] {Nara Women's University, Nara 630-8506, Japan}
\address[a:UBC] {University of British Columbia, Vancouver, British Columbia, V6T 1Z1, Canada}
\address[a:IPP] {Institute of Particle Physics (Canada), Victoria, British Columbia V8W 2Y2, Canada}
\address[a:BINP]{Budker Institute of Nuclear Physics SB RAS, Novosibirsk 630090, Russian Federation}
\address[a:NSU]{Novosibirsk State University, Novosibirsk 630090, Russian Federation}
\address[a:SOKENDAI]{The Graduate University for Advanced Studies (SOKENDAI), Hayama 240-0193, Japan}
\address[a:KEK]{High Energy Accelerator Research Organization (KEK), Tsukuba 305-0801, Japan}
\address[a:PerugiaUNIV]{Dipartimento di Fisica, Universit\`{a} di Perugia, I-06123 Perugia, Italy}
\address[a:PerugiaINFN]{INFN Sezione di Perugia, I-06123 Perugia, Italy}
\address[a:Hanyang]{Department of Physics and Institute of Natural Sciences, Hanyang University, Seoul 04763, South Korea}
\address[a:Duke]{Duke University, Durham, North Carolina 27708, U.S.A.}

\author[a:uvic]{S.~Longo\corref{correspondingauthor}\fnref{fnPS}}
\ead{longos@uvic.ca}

\author[a:uvic,a:IPP]{J.M.~Roney}

\author[a:PerugiaUNIV,a:PerugiaINFN]{C.~Cecchi}

\author[a:DESY]{S.~Cunliffe}

\author[a:DESY]{T.~Ferber}

\author[a:NaraWu]{H.~Hayashii}

\author[a:UBC,a:IPP]{C.~Hearty}

\author[a:UBC]{A.~Hershenhorn}

\author[a:BINP,a:NSU]{A.~Kuzmin}

\author[a:PerugiaINFN]{E.~Manoni}

\author[a:Duke]{F.~Meier}

\author[a:NaraWu]{K.~Miyabayashi}

\author[a:KEK,a:SOKENDAI]{I.~Nakamura}

\author[a:BINP,a:NSU]{M.~Remnev}

\author[a:uvic]{A.~Sibidanov}

\author[a:Hanyang]{Y.~Unno}

\author[a:BINP,a:NSU]{Y.~Usov}

\author[a:BINP,a:NSU]{V.~Zhulanov}

\cortext[correspondingauthor]{Corresponding author}

\fntext[fnPA]{Present Address: Deutsches Elektronen--Synchrotron, 22607 Hamburg, Germany}

\begin{abstract}
This paper describes the implementation and performance of CsI(Tl) pulse shape discrimination for the Belle II electromagnetic calorimeter, representing the first application of CsI(Tl) pulse shape discrimination for particle identification at an electron-positron collider.  The pulse shape characterization algorithms applied by the Belle II calorimeter are described.  Control samples of $\gamma$, $\mu^+$, $\pi^\pm$, $K^\pm$ and $p/\bar{p}$ are used to demonstrate the significant insight  into the secondary particle composition of calorimeter clusters that is provided by CsI(Tl) pulse shape discrimination.  Comparisons with simulation are presented and provide further validation for newly developed CsI(Tl) scintillation response simulation techniques, which when incorporated with GEANT4 simulations allow the particle dependent scintillation response of CsI(Tl) to be modelled. Comparisons between data and simulation also demonstrate that pulse shape discrimination can be a new tool to identify sources of improvement in the simulation of hadronic interactions in materials.  The $K_L^0$ efficiency and photon-as-hadron fake-rate of a multivariate classifier that is trained to use pulse shape discrimination is presented and comparisons are made to a shower-shape based approach.   CsI(Tl) pulse shape discrimination is shown to reduce the photon-as-hadron fake-rate by over a factor of 3 at photon energies of 0.2 GeV and over a factor 10 at photon energies of 1 GeV.
\end{abstract}

\begin{keyword}
Belle II,
Calorimeters,
Pulse Shape Discrimination,
Thallium doped Cesium Iodide,
Particle Identification,
GEANT4
\end{keyword}

\end{frontmatter}

\newcommand{\pp}{$p$}
\newcommand{\pio}{$\pi^0$}
\newcommand{\phiCalc}{$\phi$}
\newcommand{\ap}{$\bar{p}$}
\newcommand{\kn}{$K^-$}
\newcommand{\kp}{$K^+$}
\newcommand{\kl}{$K_L^0$}
\newcommand{\kspipi}{$K_S^0\rightarrow \pi^+ \pi^-$}
\newcommand{\csi}{CsI(Tl)}
\newcommand{\dedx}{$\text{dE}/\text{d}x$}
\newcommand{\plab}{$p_\text{lab}$}
\newcommand{\eeto}{$e^+ e^- \rightarrow \,$}
\newcommand{\mmg}{$\mu^+ \mu^- (\gamma)$}
\newcommand{\kaon}{${K}^0$}
\newcommand{\kbar}{$\bar{K}^0$}
\newcommand{\belleII}{Belle~II}
\newcommand{\eoen}{$\text{E}_1\text{E}_9$}


\section{Introduction}

The \belleII{} experiment at the SuperKEKB asymmetric electron-positron collider plans to integrate a $50 \text{ ab}^{-1}$ dataset while operating at and near the $\Upsilon(4S)$ resonance, corresponding to a centre-of-mass energy of $10.58$ GeV.  With this unprecedented dataset, Belle II will search for physics beyond the Standard Model through precision flavour sector measurements and searches for rare/forbidden processes as well as darks sectors \cite{BelleIIPhysicsBook,BelleIIZprime}.  Complimentary to the improved statistical precision to be achieved at \belleII{}, advancements in new experimental techniques can also further enhance \belleII{} sensitivities to new physics and potentially allow for new measurements.  This paper presents the performance of a new method for calorimeter-based particle identification in high energy physics through the novel application of pulse shape discrimination (PSD) with thallium-doped cesium iodide (\csi{}) scintillator crystals.  The results presented demonstrate that with pulse shape discrimination direct insight into the secondary particle composition of calorimeter clusters is gained, providing a means for highly effective discrimination between electromagnetic and hadronic showers in the \belleII{} calorimeter.  This information is shown to be independent of that from other particle identification observables in \belleII{}, including current calorimeter-based quantities, leading to improvements in \kl{} vs. photon identification.  The results presented represent the first application of this experimental technique at a $e^+ e^-$ collider experiment and opens the way to improve many \belleII{} measurements where hadronic shower identification is crucial, such as in the flag ship measurement of $\sin 2 \phi_1$ using $B \rightarrow J/\psi K^0_L$ \cite{BelleIIPhysicsBook}.

The scintillation response of \csi{} is empirically well known to depend on the ionization \dedx{} of the particle that is depositing energy in the crystal \cite{Storey}.  Highly ionizing particles, such as stopping protons or alpha particles, are observed to produce \csi{} scintillation emission with a faster decay time relative to the \csi{} scintillation emission produced from energy deposits by photons or low \dedx{} particles \cite{Storey, Longo_2018,Voss2014,Skulski}.  This phenomenon has been widely exploited in the energy regime of $<$ 10 MeV to discriminate between interactions caused by electrons, protons and alpha particles \cite{Storey,Skulski}, as well as in nuclear physics for low energy nuclei identification \cite{Voss2014,AMPHORA,CHIMERA,Bendel2013}.   Recent studies have demonstrated the significant potential for PSD to improve hadronic shower identification at electron-positron collider experiments \cite{Longo_2018,Bartle,McLean2006,Ashida2018}.   Although several past and present detectors operating at high energy $e^+ e^-$ colliders have employed \csi{} calorimeters, such as Belle \cite{belle2002}, BaBar \cite{babar2002,babar2013} and BESIII \cite{besiii2009}, applying \csi{} PSD to improve particle identification has yet to be attempted at a high energy $e^+ e^-$ collider experiment.  

This paper presents the implementation and performance of \csi{} pulse shape discrimination with the \belleII{} calorimeter using collision data collected during the summer 2018 commissioning of the \belleII{} experiment.  The data used in this analysis corresponds to an integrated luminosity of $0.5 \text{ fb}^{-1}$ collected at the $\Upsilon(4S)$ resonance \cite{BelleIILumi}.  

The \belleII{} detector has a cylindrical geometry and is constructed from of a collection of sub-detectors that together perform as a spectrometer operating in a 1.5 T magnetic field.  The innermost sub-detector is a vertex detector beginning at a radius of 14 mm from the interaction point and consists of two layers of pixel detectors followed by four layers of double-sided silicon strip detectors.  During the summer 2018 commissioning of the \belleII{} experiment only one octant of the vertex detector was installed.  Extending from a radius of 160 mm to 1100 mm is the central drift chamber, which applies a 50\% He, 50\% $\text{C}_2 \text{H}_6$ gas mixture to perform charged particle detection, and identification through dE/d$x$ measurements.  After the drift chamber is a charged particle identification system consisting of a Cherenkov-based time-of-propagation detector in the barrel region and an aerogel ring imaging Cherenkov detector in the forward region.  The electromagnetic calorimeter is constructed from \csi{} scintillator crystals and includes a barrel region beginning at a radius of 1250 mm, in addition to forward and backward endcaps.  The outermost sub-detector is the \kl{}-muon detector system.  The endcaps and the initial layers of the barrel region of the \kl{}-muon detector are constructed from alternating layers of iron plates and scintillating strip detectors.  The outer barrel layers substitute the scintillating strip detectors with resistive-plate chambers. Additional details of the \belleII{} detector can be found in reference \cite{BelleIITDR}. 

This paper is organized as follows:  Section \ref{sec_CalorDes} describes the experimental details concerning the implementation of \csi{} PSD with the \belleII{} electromagnetic calorimeter.  In Section \ref{sec_controlSamples} the pulse shapes of crystals in calorimeter clusters produced by control samples of $\gamma$, $\mu^+$, $\pi^\pm$, $K^\pm$ and $p/\bar{p}$ selected from \belleII{} data and simulation are studied.   This survey demonstrates that by analysing the scintillation pulse shapes of the \csi{} crystals in a calorimeter cluster,  the types of secondary particles produced in the cluster can be identified.  In Section \ref{sec_PIDwPSD} the performance of a multivariate classifier, which is trained to use \csi{}  PSD to separate \kl{} and photons, is measured and compared with a shower-shape based approach to neutral particle identification.  Section \ref{conclustions} presents the conclusions of this study and discusses areas for further development of this new experimental technique.

\section{Pulse Shape Discrimination with the \belleII{} Calorimeter}
\label{sec_CalorDes}

In this section the reconstruction and simulation methods implemented to apply pulse shape discrimination with the \belleII{} calorimeter are described.  The relevant features of the \belleII{} calorimeter signal chain that allow for \csi{} waveforms to be digitized and recorded for offline pulse shape analysis are outlined.  The waveform shape characterization techniques are then described and the methods applied to simulate the ionization-dependent \csi{} response are discussed.

The \belleII{} calorimeter re-uses the calorimeter of Belle, but with upgraded electronics following the initial pre-amplification stage.  The calorimeter is constructed from 8736 CsI(Tl) scintillator crystals that have a trapezoidal geometry with front face area of $\sim4.5 \times 4.5 \text{ cm}^2$, rear face area of $\sim5 \times 5 \text{ cm}^2$ and nominal length of 30 cm.  Each crystal is equipped with two Hamamatsu S2744-08 photodiodes, which have a surface area of $10 \times 20 \text{ mm}^2$ and are glued to the rear crystal face \cite{BelleIITDR}.  Two pre-amplifiers, one for each photodiode, are also mounted on the rear of the crystal to integrate the signal emitted by each photodiode \cite{BelleIITDR}.
  
Following the initial pre-amplification stage, the two signals are summed then processed by a $\text{CR-(RC)}^4$ shaping amplifier with shaping time of $0.5$ $ \mu$s \cite{EMCalBelleII}.  The signal is then digitized into 31 samples with 18-bit precision and at a sample frequency of 1.7669 MHz (sample time of 0.56594 $\mu$s).  During data-taking the digitized waveform is processed online with Field-Programmable-Gate-Array's (FPGA's) to measure the magnitude and time of the energy deposit in the crystal.  At present, the waveform analysis by the FPGA's is limited to computing only the energy and time of the waveform.  As this does not explicitly contain information that characterizes the waveform shape, which is required for PSD, an upgrade of the FPGA firmware was implemented such that if the energy measurement by the FPGA exceeds 30 MeV then the 31 waveform data points are stored offline. Although crystals with energy deposits below 30 MeV are expected to contain PSD information, the 30 MeV energy threshold is applied due to the bandwidth limitations of the \belleII{}  data acquisition system, which is unable to record the waveforms of all 8736 CsI(Tl) channels for every event.  Due to energy deposits from beam backgrounds produced by SuperKEKB, the number of crystals per event which are above a given energy threshold grows rapidly as the threshold decreases below 30 MeV.  To maximize the PSD performance, the 30 MeV value for this threshold was determined to be the minimal value that the data acquisition system could sustain given the beam backgrounds levels experienced during operation \cite{Longo_thesis}.

To characterize the waveform pulse shape the techniques developed in reference \cite{Longo_2018} are applied.  The study in reference \cite{Longo_2018} demonstrated that energy deposits by highly ionizing particles produce a CsI(Tl) scintillation component measured to have a decay time of $630\pm10$ ns.  This scintillation component is referred to as the \textit{hadron component} as it is only produced by highly ionizing energy deposits and thus not present in scintillation emission from electromagnetic showers or energy deposits from low \dedx{} particles, such as minimum ionizing particles \cite{Longo_2018}.  The shape of the CsI(Tl) waveform is characterized by the crystal \textit{hadron intensity} defined as the fraction of scintillation emission emitted in the hadron component relative to the total scintillation emission.  

To measure the magnitude of the total and hadron component scintillation emission, the waveforms recorded offline are fit to the model defined in equation \ref{PhotonHadronFunction}.

  \begin{equation}
G(t) =  L_\text{Photon} R_\text{Photon}(t-t_0) + L_\text{Hadron} R_\text{Hadron}(t-t_0)
\label{PhotonHadronFunction}
\end{equation} 

\noindent In equation \ref{PhotonHadronFunction}, 

\begin{flushleft}
 $G(t)$ is the \csi{} waveform. \\~\\
 
 $t$ is time. \\~\\
 
 $t_0$ is the time the incident particle's energy is deposited. \\~\\

$R_\text{Photon}$ is the photon template, which is defined as the shape of the signal at the output of the full signal chain of a \belleII{} crystal channel associated with the \csi{} scintillation produced in an electromagnetic shower. \\~\\

$R_\text{Hadron}$ is the hadron template, which  is defined as the shape of the signal at the output of the full signal chain of a \belleII{} crystal channel associated with the additional \csi{} scintillation component produced when hadronic interactions take place.\\~\\
 
$L_\text{Photon}$ is the photon scintillation component light output yield.\\~\\
 
$L_\text{Hadron}$ is the hadron scintillation component light output yield.\\~\\
\end{flushleft}

From the quantities measured by the model described in equation \ref{PhotonHadronFunction}, the crystal energy, $\text{E}^\text{crystal}_\text{Total}$, and crystal hadron intensity are computed using equations  \ref{crystalEnergy} and \ref{crystalHadronIntensity}, respectively.

 \begin{equation}
\label{crystalEnergy}
\text{E}^\text{crystal}_\text{Total} =  L_\text{Photon} +   L_\text{Hadron}
 \end{equation} 

 \begin{equation}
\label{crystalHadronIntensity}
\text{Hadron Intensity} = \frac{L_\text{Hadron}}{L_\text{Photon} +   L_\text{Hadron}}
 \end{equation} 

From these definitions, electromagnetic shower energy deposits are expected to have hadron intensity of zero, whereas energy deposits from highly ionizing particles, are expected to have hadron intensity greater than zero.  The exact value of the hadron intensity will depend on the magnitude of energy deposited at an ionization \dedx{} that is above the threshold required to produce hadron scintillation component emission.

\begin{figure}[!ht]
\centering

\begin{subfigure}[t]{0.49\textwidth}
  \includegraphics[width=1\linewidth]{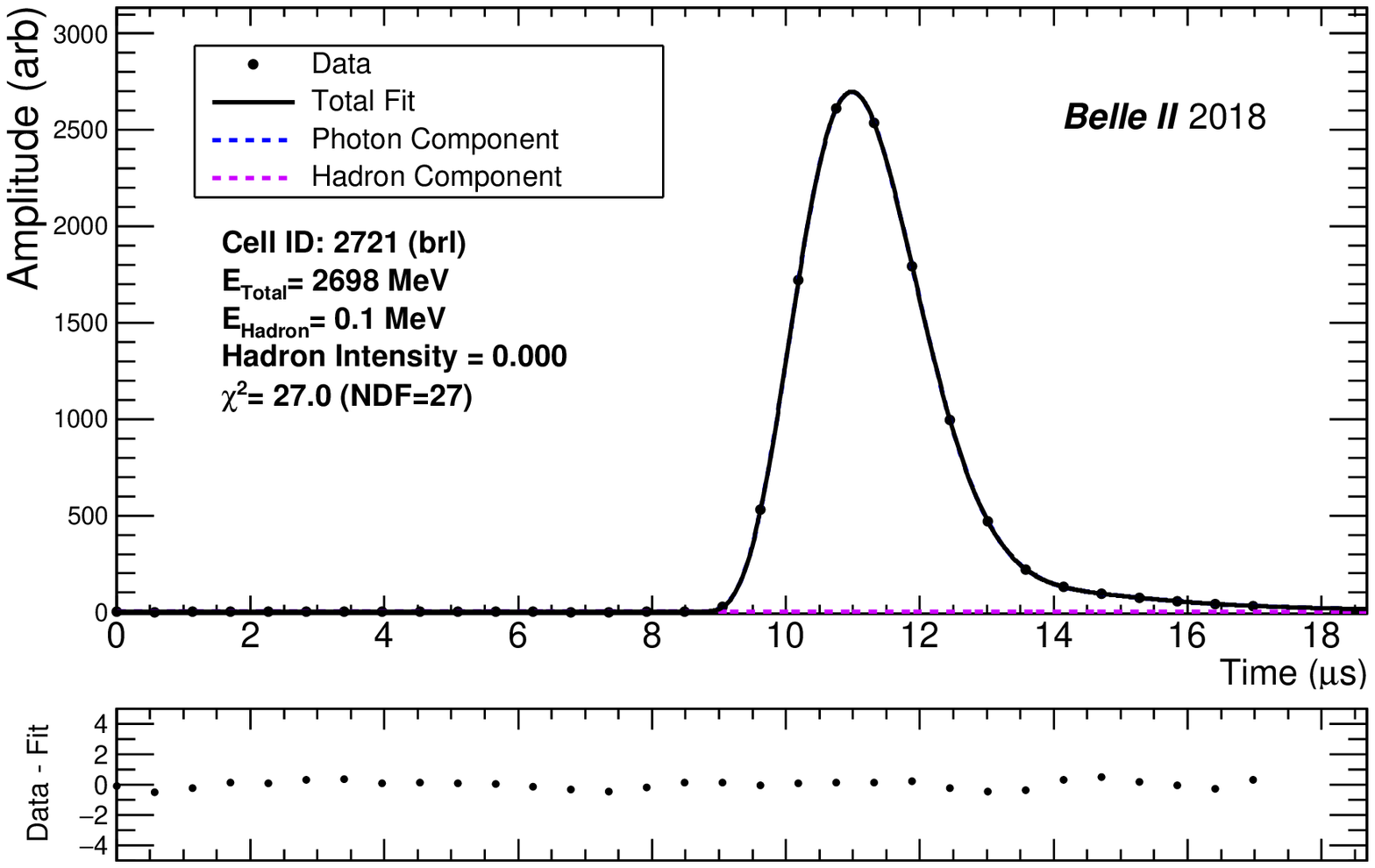}
      \caption{}
            \label{f_SampleWaveforms_a}
\end{subfigure}
\begin{subfigure}[t]{0.49\textwidth}
  \includegraphics[width=1\linewidth]{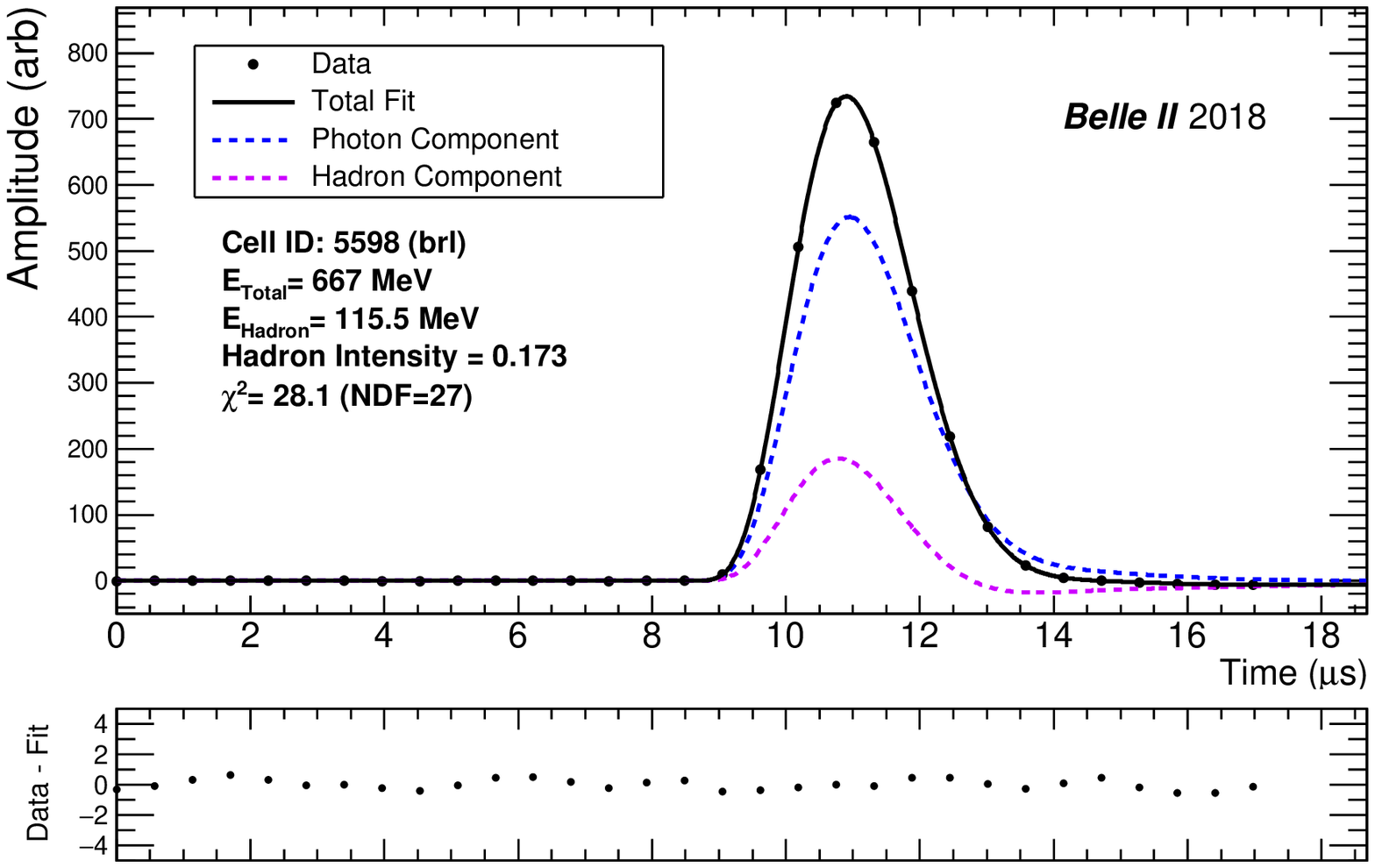}
      \caption{}
      \label{f_SampleWaveforms_b}
\end{subfigure}

\caption{Typical \belleII{} \csi{} waveforms from data with fit to Photon+Hadron templates overlaid.  a) Waveform with photon-like pulse shape, as expected from an electromagnetic shower.  b) Waveform with hadron-like pulse shape, as expected from a hadronic shower.}
\label{f_SampleWaveforms}

\end{figure}

Two examples of typical waveforms recorded during summer 2018 \belleII{} commissioning runs are shown in Figure \ref{f_SampleWaveforms}.  Figure \ref{f_SampleWaveforms_a} shows a typical waveform with a photon-like pulse shape and Figure \ref{f_SampleWaveforms_b} shows a typical waveform with hadron-like pulse shape.  Comparing these two waveforms, it is observed that hadron-like pulse shapes have a suppressed tail relative to the photon-like pulse shapes, which is well modelled by the hadron template.  

To simulate particle interactions in the \belleII{} detector, Monte Carlo (MC) simulations using GEANT4 particle interactions in matter simulation libraries are applied \cite{GEANT4}.  The GEANT4 physics list used is the \verb|FTFP_BERT|.   By default, GEANT4 does not include simulations of the ionization \dedx{} dependent \csi{} scintillation response.  To simulate the \csi{} scintillation response to highly ionizing particles we apply the simulation methods described in reference \cite{Longo_2018}.  These methods compute the magnitude of the hadron scintillation component emission and Birk's scintillation efficiency \cite{BirksTheoryandPractice} using the instantaneous \dedx{} of the primary and secondary particles that contribute to the total energy deposited in the \csi{} crystal. Simulated waveforms are constructed by iterating over all discrete energy deposits in the crystal volume and accumulating a template sum, weighted by the corresponding scintillation light output contributions from photon component scintillation emission and hadron component scintillation emission.  To model the detector noise conditions, noise waveforms recorded from events that are randomly triggered during data-taking runs are added to the simulated waveform.  After pulse construction, the simulated pulse is fit using the same methods as described above for data.

\section{CsI(Tl) Pulse Shapes of Crystals in Clusters from a Selection of Particle Control Samples}
\label{sec_controlSamples}

This section presents a survey of the scintillation pulse shapes observed in crystals from calorimeter clusters produced by control samples of $\gamma$, $\mu^+$, $\pi^\pm$, $K^\pm$ and $p/\bar{p}$ selected from \belleII{} commissioning data.  These control samples are presented to demonstrate the variety of pulse shape signatures, which arise due to the different material interactions initiated by different types of particles, and the ability to accurately simulate them. 

Demonstrated in reference \cite{Longo_2018}, energy deposits by highly ionizing particles generate hadron scintillation component emission in \csi{}, which is differentiated from scintillation produced by low \dedx{} energy deposits due to its relatively fast scintillation time.  The \csi{} scintillation pulse shape is thus determined by the fraction of energy deposited at an ionization \dedx{} that is significant enough to produce the hadronic scintillation component emission.   This direct dependence allows the \csi{} pulse shape to be used to identify the types of secondary particles that contributed to the energy deposit in the crystal.  As a result the pulse shapes of the crystals in the cluster will depend on the primary particle type.

The main features of each control sample are summarized as:

\begin{itemize}
\item $\gamma$ - \csi{} pulse shapes of energy deposits from electromagnetic showers.
\item $\mu^\pm$ - \csi{} pulse shapes of energy deposits from minimum ionizing particles.
\item $\pi^\pm$ - \csi{} pulse shapes of crystals in hadronic showers.
\item $K^\pm$ - Due to strangeness conservation in strong interactions, hadronic interactions of $K^+$ in the momentum regime studied are suppressed relative to $K^-$ \cite{mesonInter}.  This effect is observed with \csi{} PSD, illustrating that \csi{} PSD can measure the hadronic activity in a cluster.
\item  $p/\bar{p}$ - In the momentum regime studied $p$ frequently will ionize and stop in the calorimeter.  During this process the $p$ becomes highly ionizing, demonstrating the scenario when the primary particle can directly produce hadron scintillation light output.  The pulse shape distribution is shown to be distinct from a $\bar{p}$, which annihilates in the \csi{}.  How PSD can be used to improve the GEANT4 simulation of $\bar{p}$ interactions is also discussed.
\end{itemize}

\noindent Selections for each control sample are described the respective section below.   A momentum dependent efficiency correction is applied to simulation to account for inefficiencies in data due to event triggering and reconstruction inefficiencies.  After the momentum dependent efficiency correction, the momentum distributions in data and simulation are in agreement thus allowing the calorimeter quantities to be compared  \cite{Longo_thesis}.  Only clusters in the barrel region of the calorimeter are studied.

\subsection{Photons}
\label{photoncrystals}
A sample of electromagnetic showers produced by photons with lab momentum magnitude, \plab{}, in the range  0.5 $<$ \plab{} $\leq 1$ GeV/c was selected from $e^+ e^- \rightarrow \mu^+ \mu^- (\gamma)$ events.  This selection requires the event to have two well reconstructed oppositely charged tracks in addition to a photon.  Bhabha events are rejected by requiring the calorimeter cluster energy of each track to be consistent with an ionization cluster. The mass of the $\mu^+ \mu^- \gamma$ system is required to be consistent with the total centre-of-mass energy of SuperKEKB.  In addition, the magnitude and direction of the photon momentum vector is required to be consistent with the recoil momentum of the $\mu^+ \mu^-$ system.  Backgrounds from $e^+ e^- \rightarrow \pi^+ \pi^- (\gamma)$ and $e^+ e^- \rightarrow K^+ K^- (\gamma)$ are suppressed by vetoing events where the $\mu^+ \mu^-$ invariant mass is consistent with a $\rho$ or $\phi$ \cite{Longo_thesis}.

In this momentum range a photon is likely to interact in the \csi{ calorimeter by generating an electromagnetic shower, consisting of only secondary electrons, positrons and photons.  Due to the absence of highly ionizing particles in the electromagnetic shower,  the hadron component intensity values of the crystals in the photon clusters are expected to be distributed close to zero, independent of the photon energy and the crystal energy. 

Shown in Figure \ref{f_photonPulseShapes} is the crystal hadron intensity vs. crystal energy distribution for the crystals in the selected photon clusters.  Observed in this figure, the crystal hadron intensity values are distributed near zero in data and simulation, as expected.  The hadron intensity fluctuations about zero present in these distributions arise because a small hadron component contribution can artificially be added during the multi-component fit to compensate for noise present in the waveform.  As observed in the data and simulation, this fitting effect results in the hadron intensity to sometimes have small negative values despite the true hadron intensity always being greater or equal to zero.  

\begin{figure}[!ht]
\centering

\begin{subfigure}[t]{.49\textwidth}
  \includegraphics[width=1\linewidth]{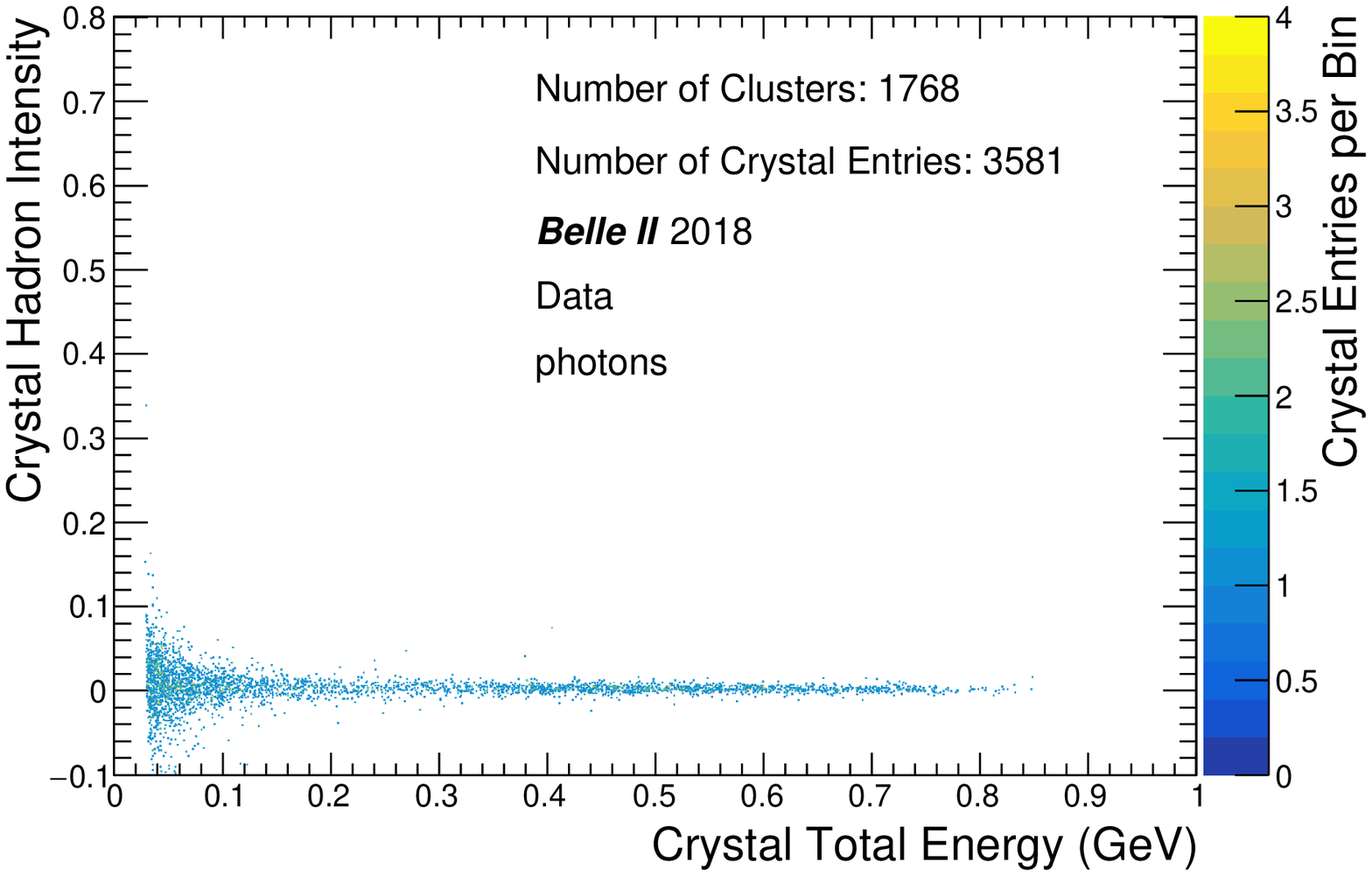}
        \caption{}
        \end{subfigure}
\begin{subfigure}[t]{.49\textwidth}
  \includegraphics[width=1\linewidth]{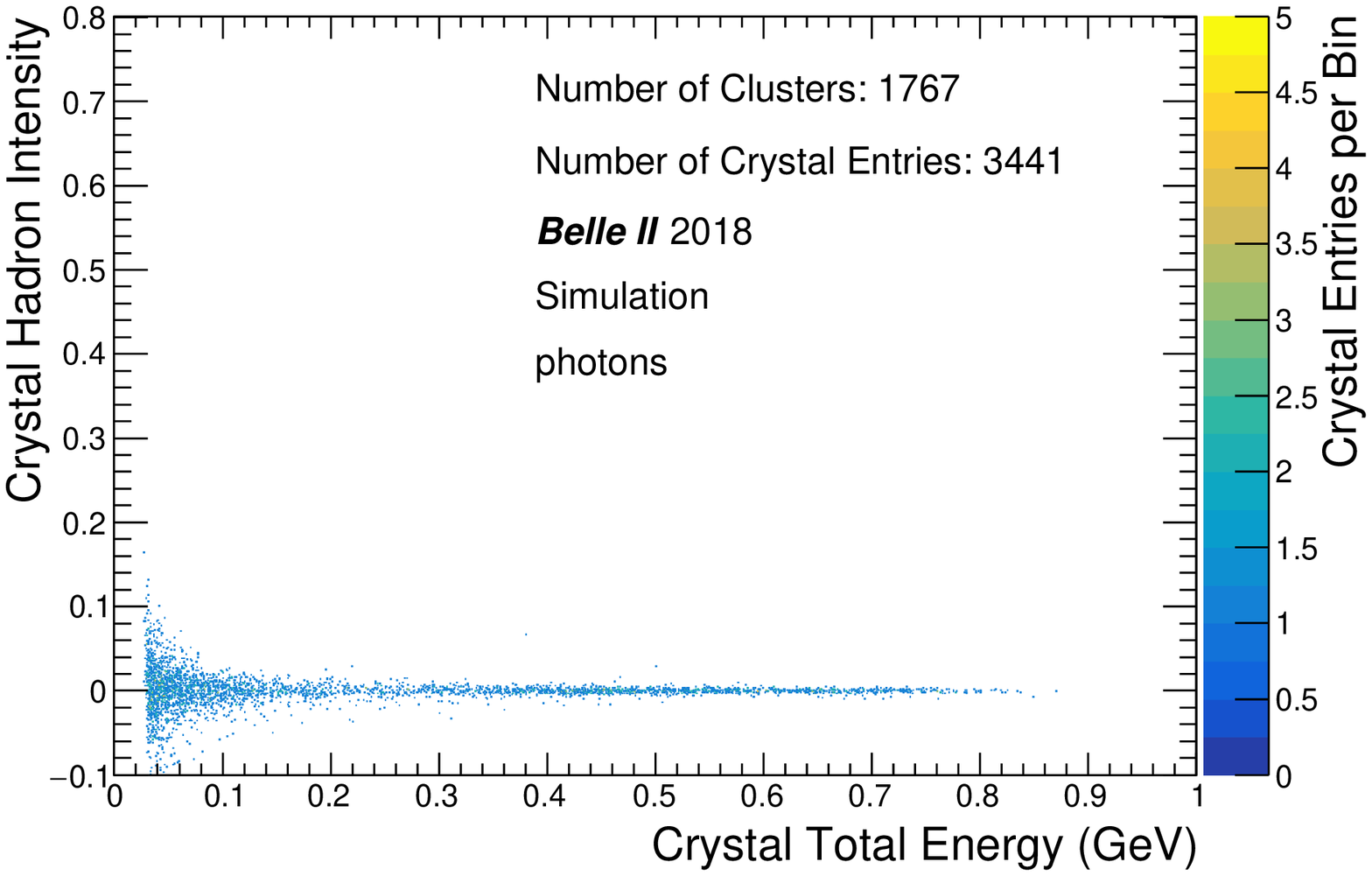}
        \caption{}
\end{subfigure}

\caption{Crystal hadron intensity vs crystal energy distributions for crystals in calorimeter clusters produced by 0.5 $<$ \plab{} $\leq 1$ GeV/c photons selected from \eeto{}\mmg{}, a) data b) simulation.}
\label{f_photonPulseShapes}

\end{figure}

\subsection{Muons}

A sample of calorimeter clusters produced by muons with momentum magnitude in the range 1 $<$ \plab{} $\leq 7$ GeV/c was selected using $e^+ e^- \rightarrow \mu^+ \mu^- (\gamma)$ events \cite{Longo_thesis}.  At this momentum scale, the dominant interaction for muons in \csi{} is ionization.  If the muon has sufficient transverse momentum, frequently it will traverse the entire 30 cm depth of the \csi{} calorimeter, resulting in a $\sim200$ MeV total energy deposit from ionization.  This energy deposit will be concentrated a compact region of 1-2 crystals.

Figure \ref{f_muonPulseShapes} shows the crystal hadron intensity vs. crystal energy distribution for crystals in the sample of calorimeter clusters produced by the selected muons.  The abundance of crystals with total energy of $\sim200$ MeV in these distributions originates from calorimeter clusters where the muon trajectory constrained the muon ionization energy deposit to be contained mostly in a single crystal.   The sample of crystals with energies below $\sim200$ MeV are caused by the clusters where the muon trajectory traverses over multiple crystals, thus the energy deposition is divided into those crystals.  The crystal energy deposits above 250 MeV are typically from electromagnetic showers generated by energetic delta rays emitted during the muon ionization.  The distributions in Figure \ref{f_muonPulseShapes} demonstrate that independent of the crystal energy the hadron intensity values of the crystals are distributed near zero, similar to the crystals in the photon control sample.  This observation is consistent with the measurements presented in reference \cite{Longo_2018}, further demonstrating the ionization \dedx{} of the muons in this sample is too low to produce significant amounts of hadron scintillation component emission.  The pulse shapes of crystals in clusters from $\mu^-$ were observed to display the same characteristics as the crystals from $\mu^+$ clusters shown in Figure \ref{f_muonPulseShapes} \cite{Longo_thesis}.

\begin{figure}[!ht]
\centering

\begin{subfigure}[t]{.49\textwidth}
  \includegraphics[width=1\linewidth]{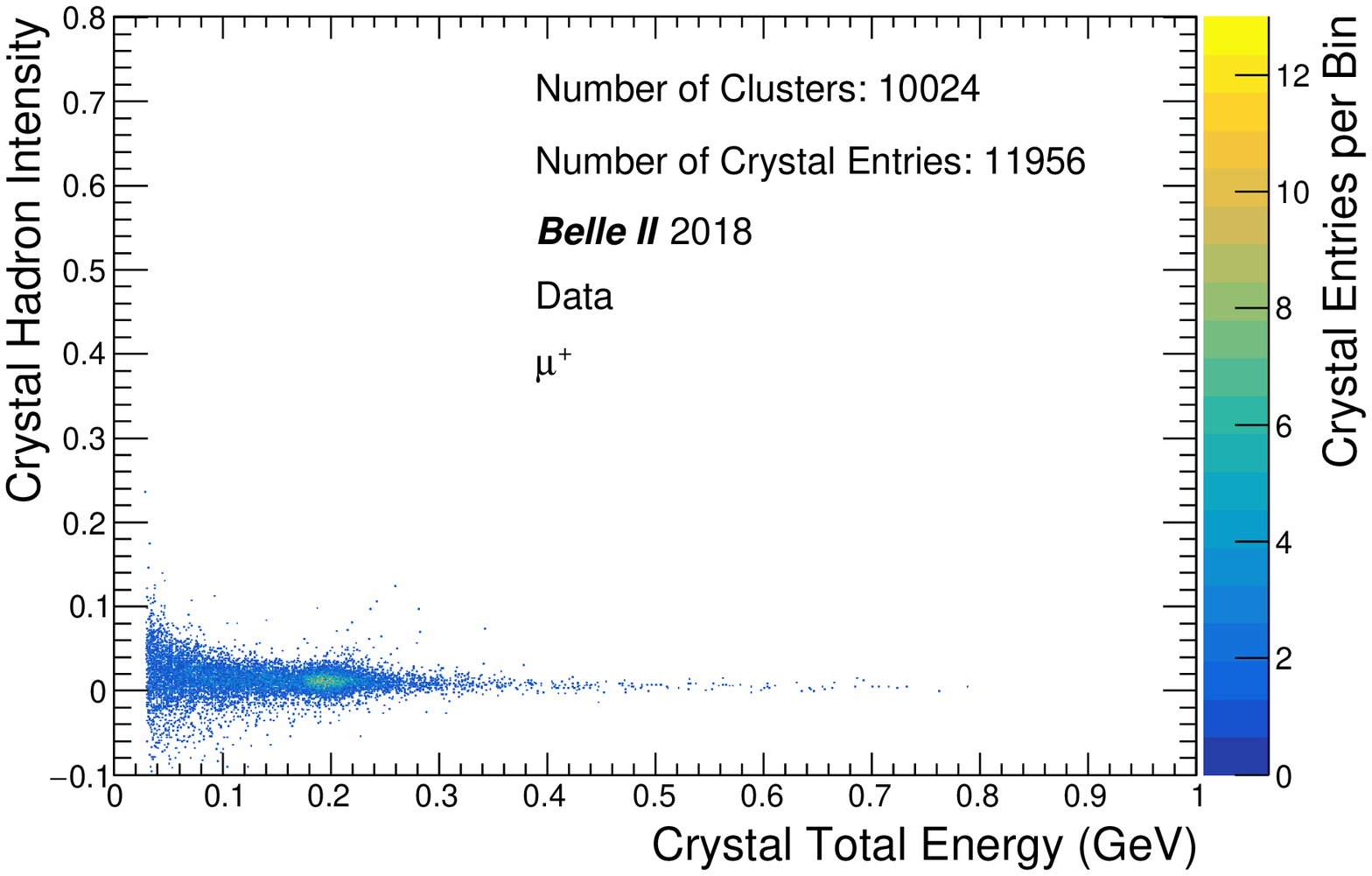}
        \caption{}
        \end{subfigure}
\begin{subfigure}[t]{.49\textwidth}
  \includegraphics[width=1\linewidth]{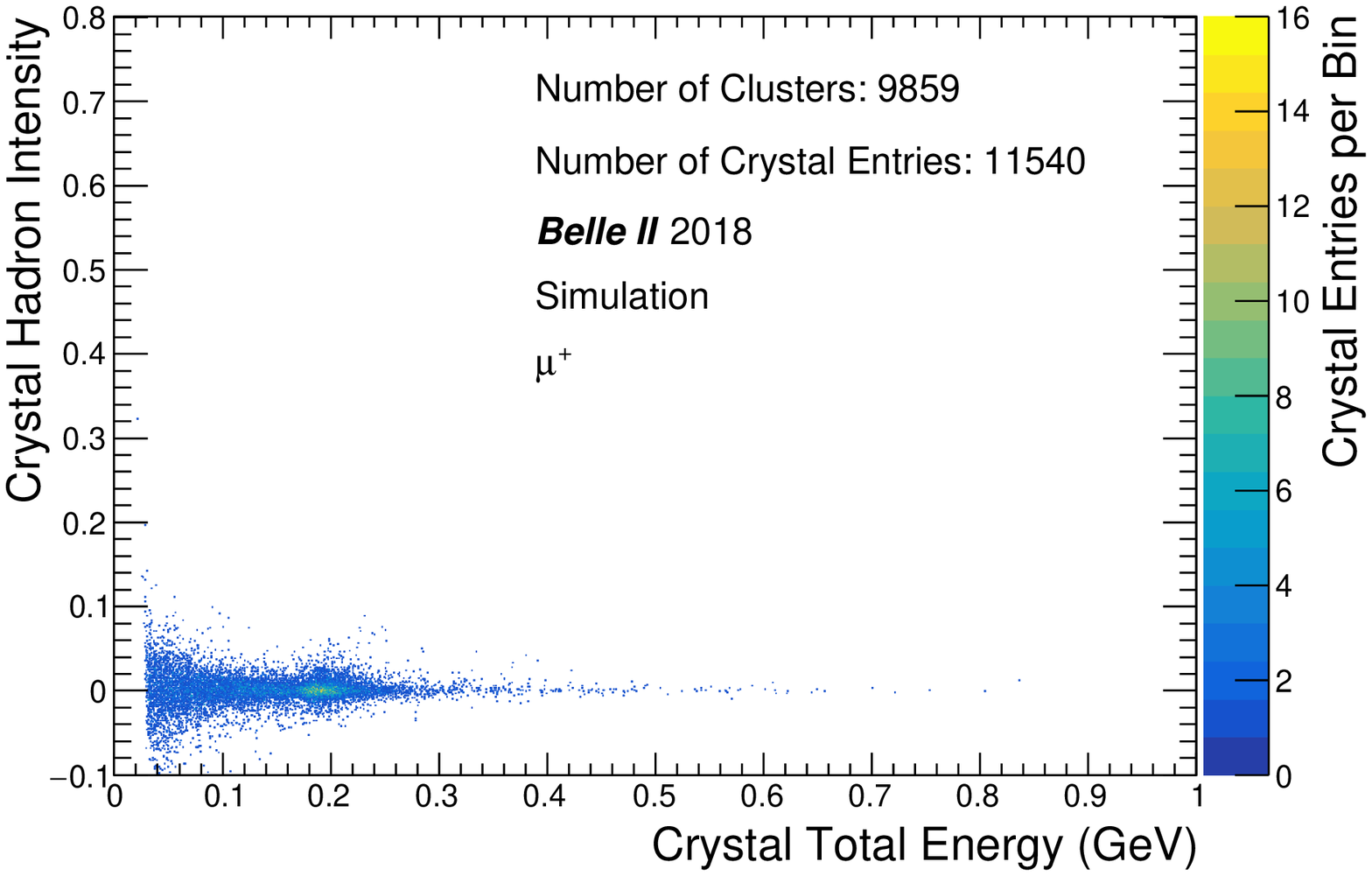}
        \caption{}
\end{subfigure}

\caption{Crystal hadron intensity vs crystal energy distributions for crystals in calorimeter clusters produced by 1 $<$ \plab{} $\leq 7$ GeV/c muons selected from \eeto{}\mmg{}, a) data b) simulation.}
\label{f_muonPulseShapes}
\end{figure}

\subsection{Charged Pions}
\label{sec_chargedpion}

A sample of charged pions in the momentum range 1 $<$ \plab{} $\leq 3$ GeV/c was selected using $K^0_S \rightarrow \pi^+ \pi^-$ decays.  This control sample provided a clean sample of charged pions by selecting tracks that formed a displaced vertex and with invariant mass consistent with the $K^0_S$ mass.   In addition, the lab momentum vector of the displaced vertex was required to be co-linear with the vector connecting the interaction point to the vertex location \cite{Longo_thesis}.  In this momentum range, a charged pion has about a 50\% probability to undergo a nuclear interaction while traversing 30 cm of \csi{} \cite{pdg}.  This results in two distinct types of calorimeter clusters, as illustrated by Figure \ref{f_pionEnergy} showing the calorimeter cluster energy distribution for the selected pion sample.  The peak in the distributions at $\sim200$ MeV corresponds to calorimeter clusters where the pion did not hadronically interact in the \csi{}.  In this case the pion leaves an ionization cluster in the calorimeter, similar to that of muons discussed in the previous section.  The remaining clusters in this distribution primarily correspond to clusters where the pion generated a hadronic shower.  

For \csi{}, the scintillation response to highly ionizing particles is known to be non-linear due to the Birk's scintillation efficiency as well as due to changes in the scintillation pulse shape \cite{Longo_2018,CALIFA}.  In Figure \ref{f_pionEnergy} two versions of simulation are overlaid to illustrate the impact that including the full \csi{} scintillation response in simulation has on the simulated pion cluster energy distributions.  In Figure \ref{f_pionEnergy} the simulation labelled \textit{No Birks and PS} uses default GEANT4 simulations, which do not include modelling of the ionization \dedx{} dependent changes in the \csi{} scintillation response. The simulation labelled \textit{w Birks and PS}, where ``PS" stands for Pulse Shape simulations, adds to the GEANT4 simulation the Birk's scintillation efficiency \cite{BirksTheoryandPractice} and pulse shape simulation techniques developed in reference \cite{Longo_2018}, which allow the ionization \dedx{} dependent \csi{} scintillation response to be modelled.  Comparing the data to the two versions of simulation, it is observed that including the full \csi{} scintillation response results in improved agreement between data and simulation.  When the full  \csi{} scintillation response is included in the simulation, a general trend observed is that the simulated cluster energies increase.  This can occur from the presence of energy deposits by secondary protons produced in the pion hadronic showers.  In the Birk's scintillation efficiency parametrization that is applied in the simulation, the \csi{} scintillation efficiency is greater than one for an intermediate \dedx{} range, as supported by data reported in references \cite{Longo_2018, CALIFA,Zeitlin2016,Koba2011,GwinMurray1963}. This results in the electron-equivalent light yield for the energy deposit from a stopping proton with kinetic energy above several MeV to be larger than the energy deposited.  This causes the measured cluster energy, which is proportional to the scintillation yield, to increase in the simulation.

\begin{figure}[!ht]
\centering

\begin{subfigure}[t]{.49\textwidth}
  \includegraphics[width=1\linewidth]{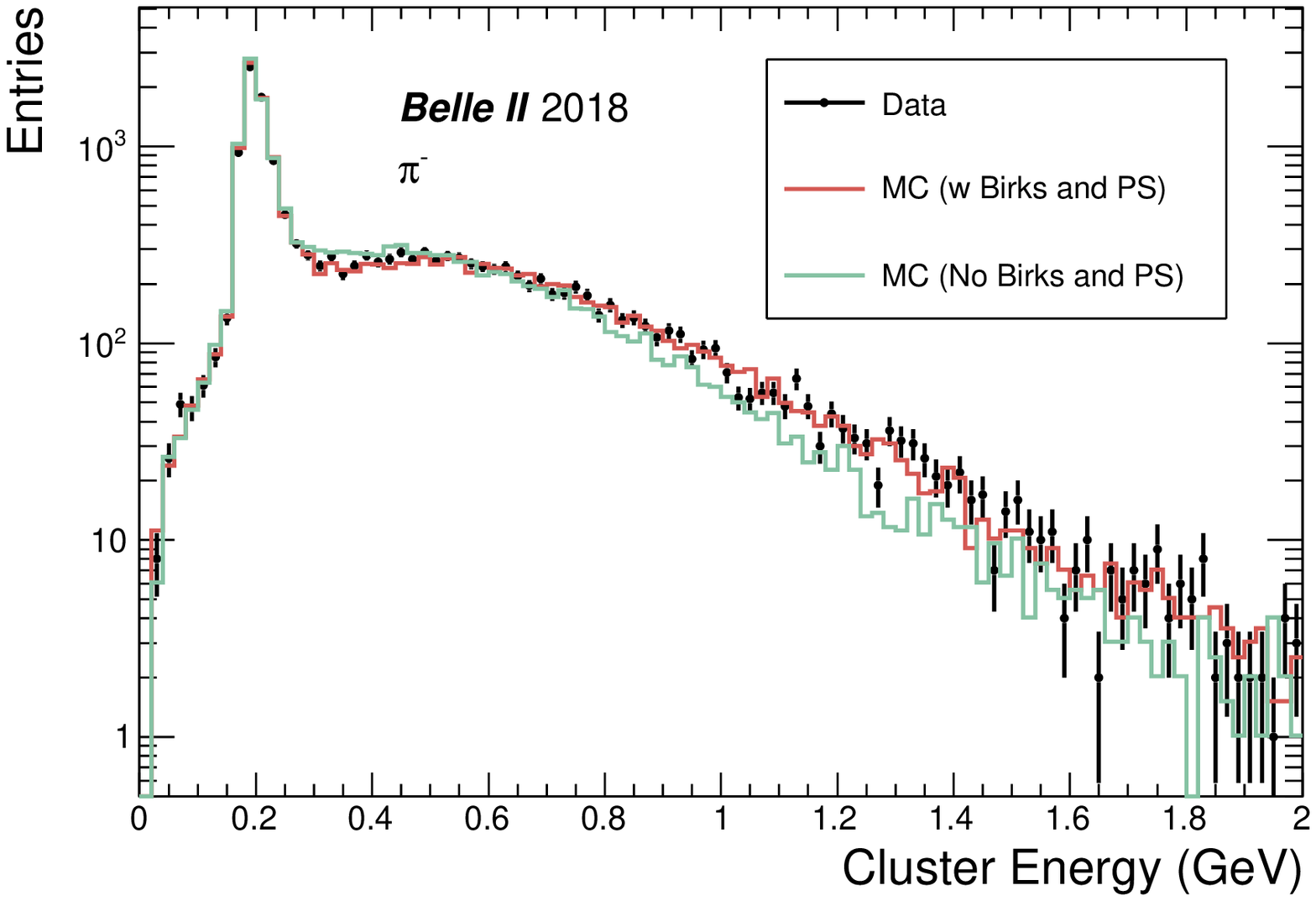}
        \caption{}
        \end{subfigure}
\begin{subfigure}[t]{.49\textwidth}
  \includegraphics[width=1\linewidth]{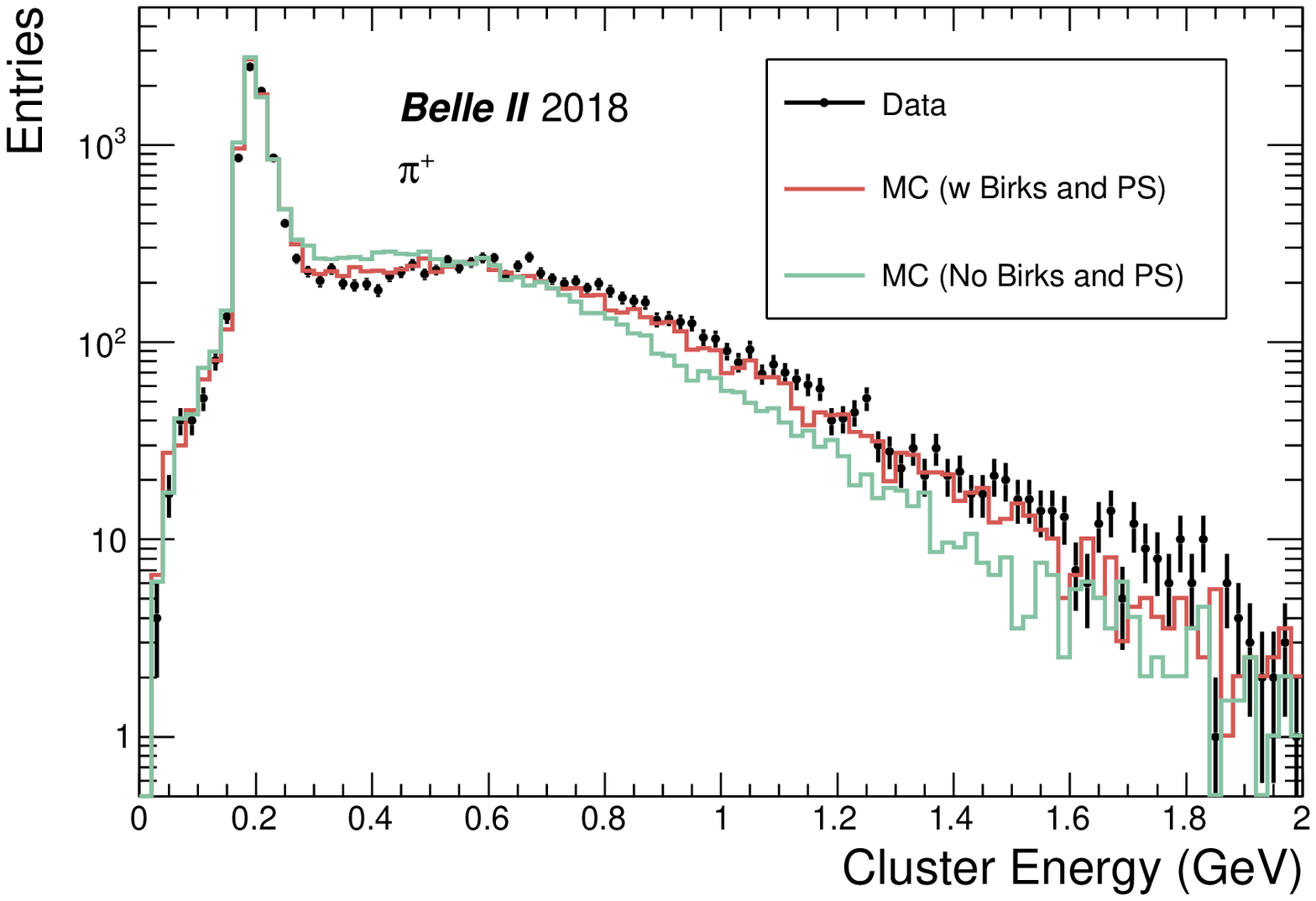}
        \caption{}
\end{subfigure}

\caption{Cluster energy distributions for charged pions a) $\pi^-$ b) $\pi^+$ in the momentum range 1 $<$ \plab{} $\leq 3$ GeV/c, selected using $K^0_S \rightarrow \pi^+ \pi^-$ decays.   The simulation labelled \textit{No Birks and PS} uses default GEANT4 simulations, which do not include modelling of the ionization \dedx{} dependent changes in the \csi{} scintillation response. The simulation labelled \textit{w Birks and PS} adds to the GEANT4 simulation the Birk's scintillation efficiency and pulse shape simulation techniques developed in reference \cite{Longo_2018}, allowing the ionization \dedx{} dependent \csi{} scintillation response to be modelled.}
\label{f_pionEnergy}

\end{figure}

The pulse shapes of crystals in pion ionization clusters were observed to have hadron intensity values distribution near zero similar to the muons shown previously.  This is attributed to the pion ionization \dedx{} being too low to produce hadronic scintillation component emission during pion ionization.  When the pion initiates a nuclear interaction however, the secondary hadrons generated in the cluster can produce significant amounts of hadron scintillation component emission.  This is illustrated by Figure \ref{f_pionPulseShapes} showing the crystal hadron intensity vs. crystal energy distribution for crystals in the 1-3 GeV/c $\pi^+$ clusters with cluster energy outside the energy range of 150-250 MeV.  This cluster energy veto removes ionization clusters, which are formed by a pion ionizing through the depth of the calorimeter without hadronically interacting, ensuring the sample is primarily composed of clusters from pion hadronic showers.  In this figure an abundance of crystals with large hadron intensity values, ranging up to 0.6, is observed in data and simulation.  This is very distinct from the photon and muon distributions discussed previously where only photon-like pulse shapes were observed and demonstrates the potential for particle identification with pulse shape discrimination.  

\begin{figure}[!ht]
\centering

\begin{subfigure}[t]{.49\textwidth}
  \includegraphics[width=1\linewidth]{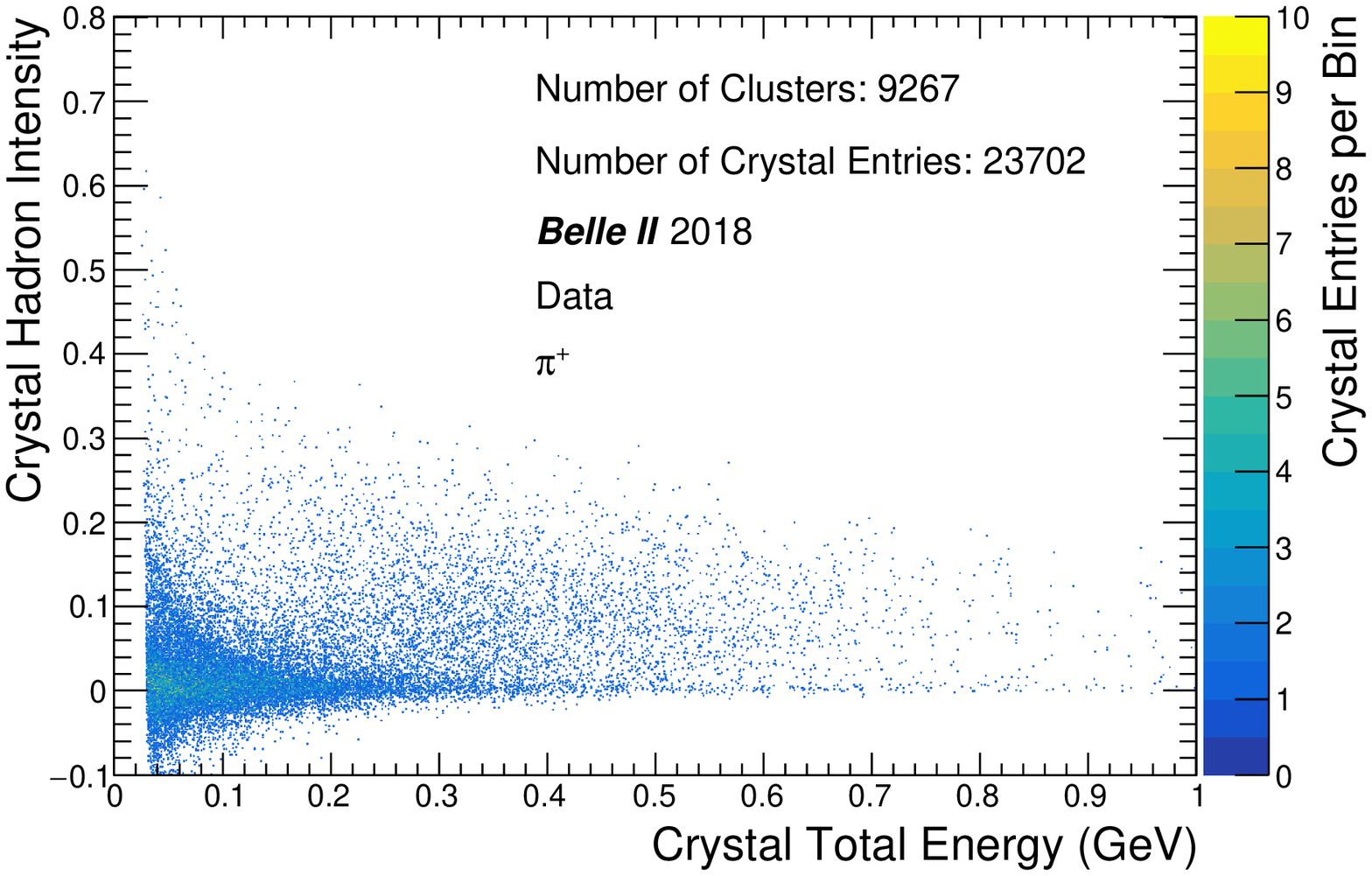}
        \caption{}
        \label{f_pionPulseShapes_dataa}
        \end{subfigure}
\begin{subfigure}[t]{.49\textwidth}
  \includegraphics[width=1\linewidth]{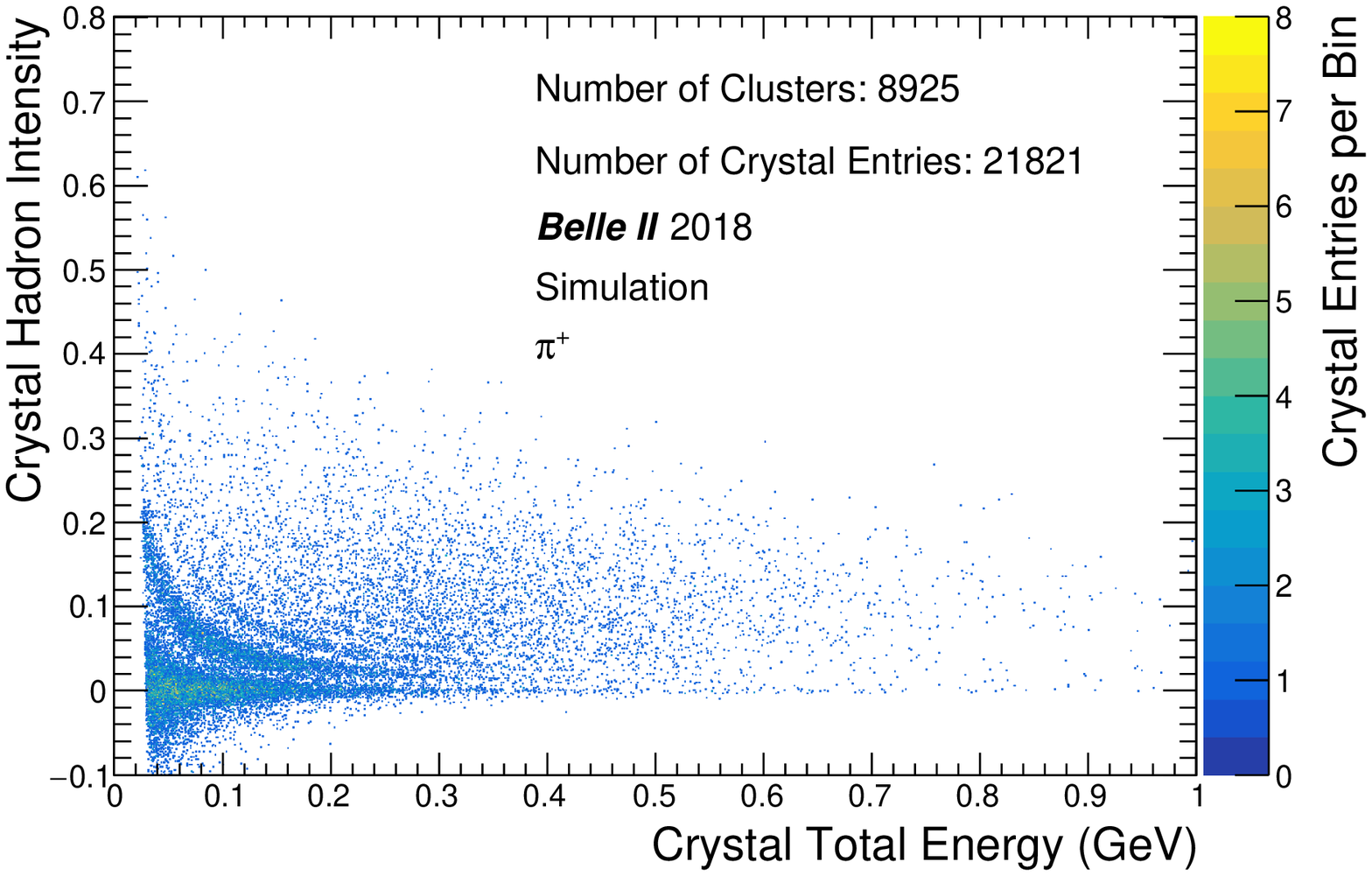}
        \caption{}
\end{subfigure}
\begin{subfigure}[t]{.49\textwidth}
  \includegraphics[width=1\linewidth]{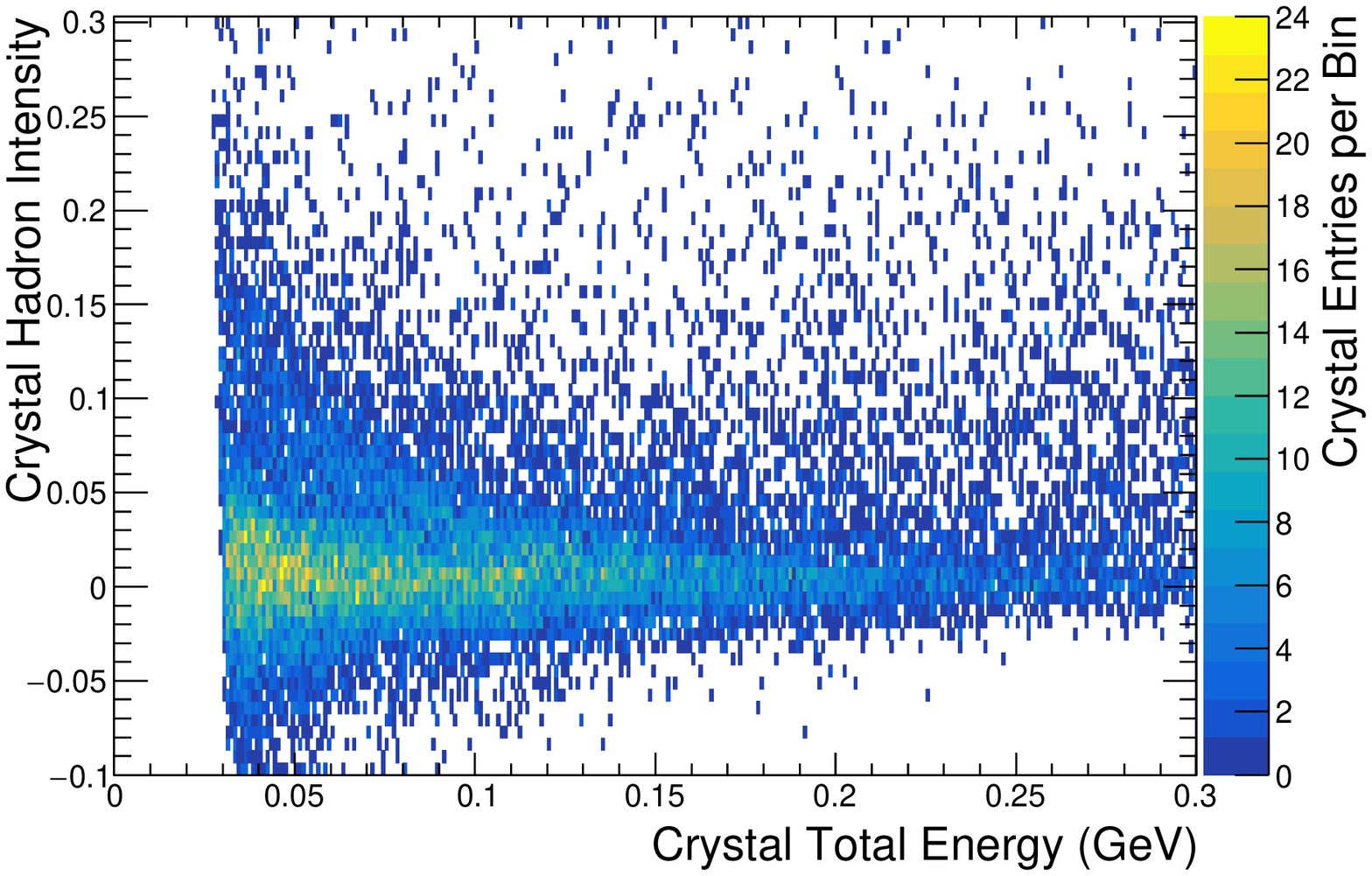}
        \caption{}
        \end{subfigure}

\caption{Crystal hadron intensity vs crystal energy distributions for crystals in calorimeter clusters produced by $\pi^+$ in the momentum range 1 $<$ \plab{} $\leq 3$ GeV/c, selected using $K^0_S \rightarrow \pi^+ \pi^-$ decays, a) data b) simulation c) zoom of data shown in Figure \ref{f_pionPulseShapes_dataa} with reduced binning allowing for the presence of the single proton band in the data to be seen in detail.}
\label{f_pionPulseShapes}

\end{figure}

The distributions shown in Figure \ref{f_pionPulseShapes} display several features, which are observed to be present in the data and simulation.  These features arise due to specific material interactions resulting in crystal energy deposits by specific compositions of secondary particles.  In the region of crystal energy below 150 MeV and hadron intensity near zero, these energy deposits are mainly from the primary pion, or a secondary charged pion emitted from the pion nuclear interaction, ionizing through a $\sim10$ cm \csi{} crystal segment before escaping the crystal volume, without initiating a nuclear interaction.  The population of crystals with energies above $\sim 300$ MeV and hadron intensity near zero however are unlikely to originate from this scenario due to the \csi{} crystal dimensions limiting the total crystal energy deposit from only pion ionization.  By investigating the GEANT4 simulation truth is was confirmed that the crystals with energy above $\sim 300$ MeV and hadron intensity near zero primarily originate from clusters where a secondary $\pi^0$ was emitted from the charged pion hadronic interaction.  The $\pi^0$ rapidly decays to two photons leaving a large energy deposit from the electromagnetic shower, which has a photon-like pulse shape.

In Figure \ref{f_pionPulseShapes} a band structure is observed in the data and simulation in the region of crystal energies below 200 MeV and hadron intensity up to 0.2.  A zoom of this feature in the data is shown in Figure \ref{f_pionPulseShapes_dataa}.  This feature is known as the single proton band \cite{Longo_2018} as crystals will have a pulse shape on this band when the crystal energy deposit is from a single proton ionizing then stopping in the crystal volume.  The protons producing these crystal energy deposits are emitted as secondary particles from the pion inelastic interaction.  As the proton kinetic energy approaches zero, the ionization \dedx{} of the proton reaches large values and significant amounts of hadron scintillation component emission are produced, resulting in a hadron-like pulse shape.   In Figure \ref{f_pionPulseShapes} the single proton band is better resolved in simulation relative to the data.  The resolution degradation observed in the data is attributed to crystal-by-crystal variations in the hadronic scintillation response, which are not included in the simulation.  As these distributions integrate information from crystals across the entire barrel section of the \belleII{} calorimeter, the accumulation of variations in hadron response, which could arise due to non-simulated factors such as differences in thallium concentration, radiation damage and diode spectral response, are expected to smear the data distribution.  The observation of the single proton band in data in Figure \ref{f_pionPulseShapes} demonstrates that crystal-by-crystal variations in the hadron scintillation response due to these factors however cannot be large, and that the hadron and photon templates are well calibrated across the calorimeter.

An additional feature observed in Figure \ref{f_pionPulseShapes} is the scatter of crystals with energy above $\sim 200$ MeV and hadron intensity above $0.02$.  Crystals in this pulse shape range are referred to as \textit{multi-hadron} pulse shapes because the energy deposits that generate these crystals are from numerous low energy hadrons (protons, neutrons, deuterons, tritons, alphas) emitted from the nucleus de-excitation, which follows the pion hadronic interaction \cite{Longo_2018}.  The significant spread in crystal energy and hadron intensity values is a result of the large variation in the energy, type and multiplicity of the secondary particles emitted from this interaction.

\subsection{Charged Kaons}

A charged kaon control sample in the lab momentum range of 0.3-0.5 GeV/c was selected using \dedx{} information measured by the \belleII{} central drift chamber \cite{Longo_thesis}.  The strangeness of kaons provides an interesting control sample to illustrate how the \csi{} pulse shapes can give a measure of the hadronic activity in a calorimeter cluster.  This is because in the momentum range 0.3-0.5 GeV/c strangeness conservation restricts the hadronic material interactions of \kp{} relative to \kn{}.  For example, a \kn{} nuclear interaction in this momentum range can produce a variety of hyperon final states such as $\pi^0 \Lambda$, $\pi^\pm \Sigma^\mp$ and $\pi^0 \Sigma^0$, which are unavailable to the \kp{} \cite{mesonInter}.    

\begin{figure}[!ht]
\centering

\begin{subfigure}[t]{.49\textwidth}
  \includegraphics[width=1\linewidth]{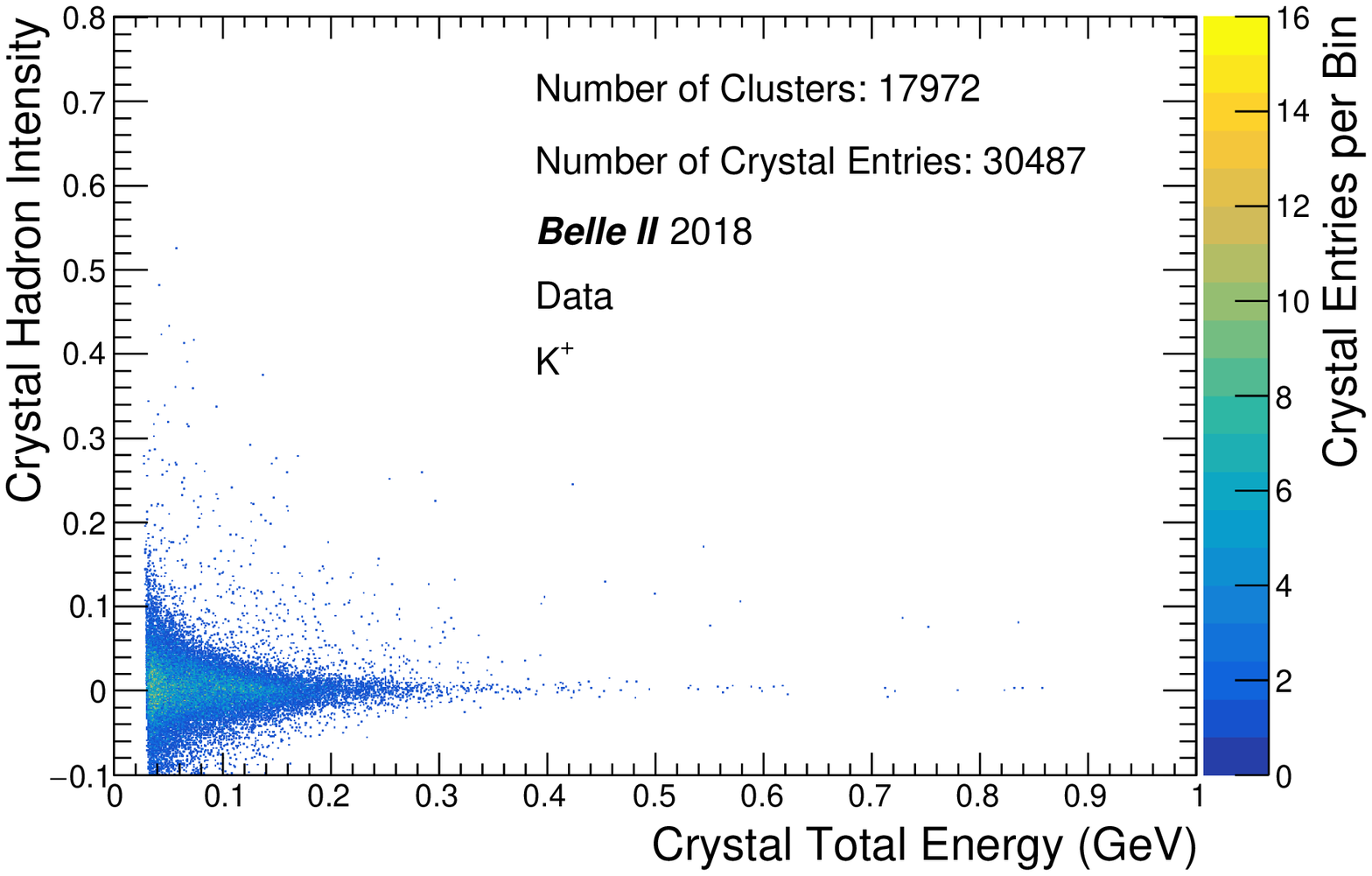}
        \caption{}
        \end{subfigure}
\begin{subfigure}[t]{.49\textwidth}
  \includegraphics[width=1\linewidth]{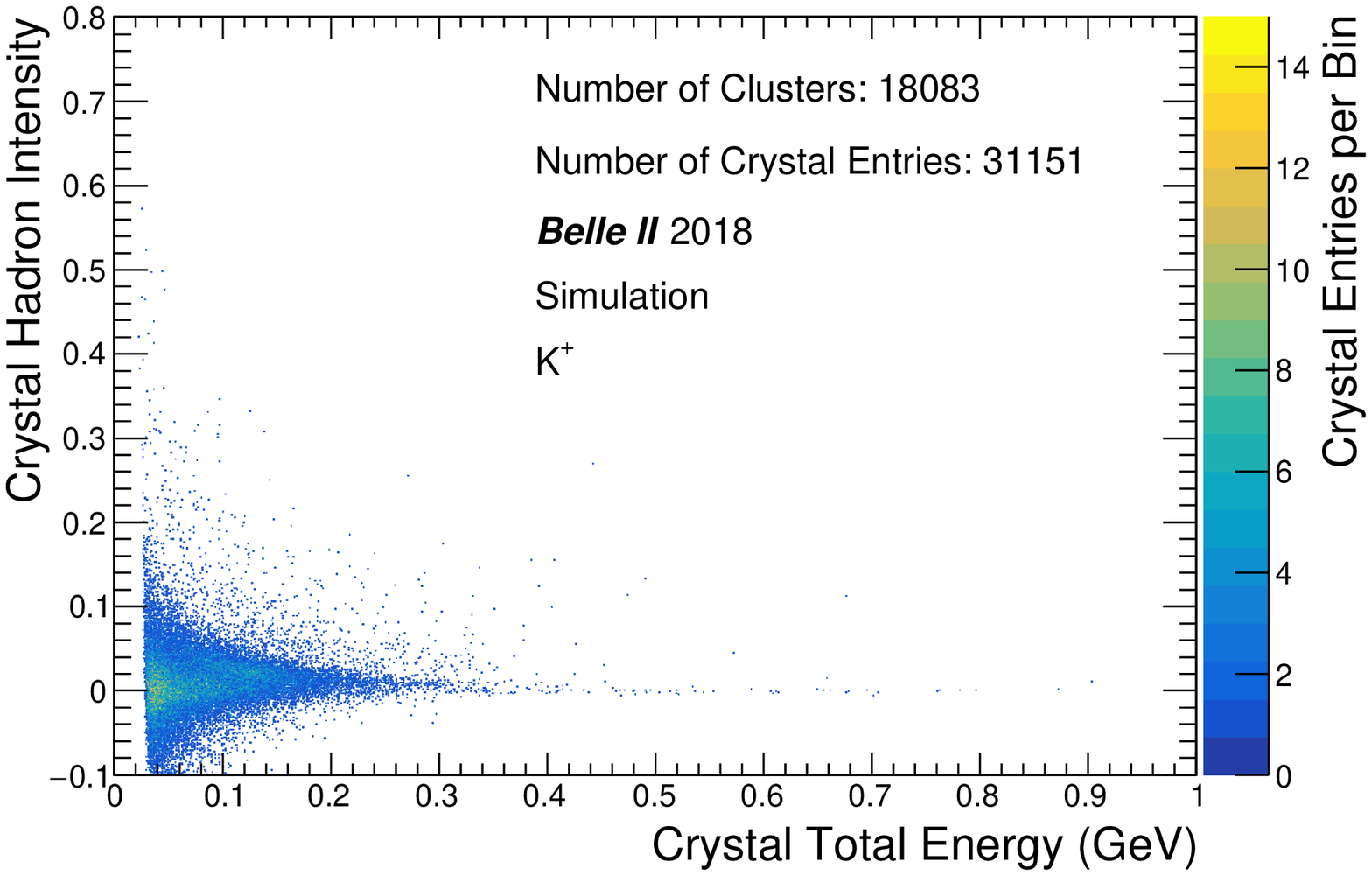}
        \caption{}
\end{subfigure}

\begin{subfigure}[t]{.49\textwidth}
  \includegraphics[width=1\linewidth]{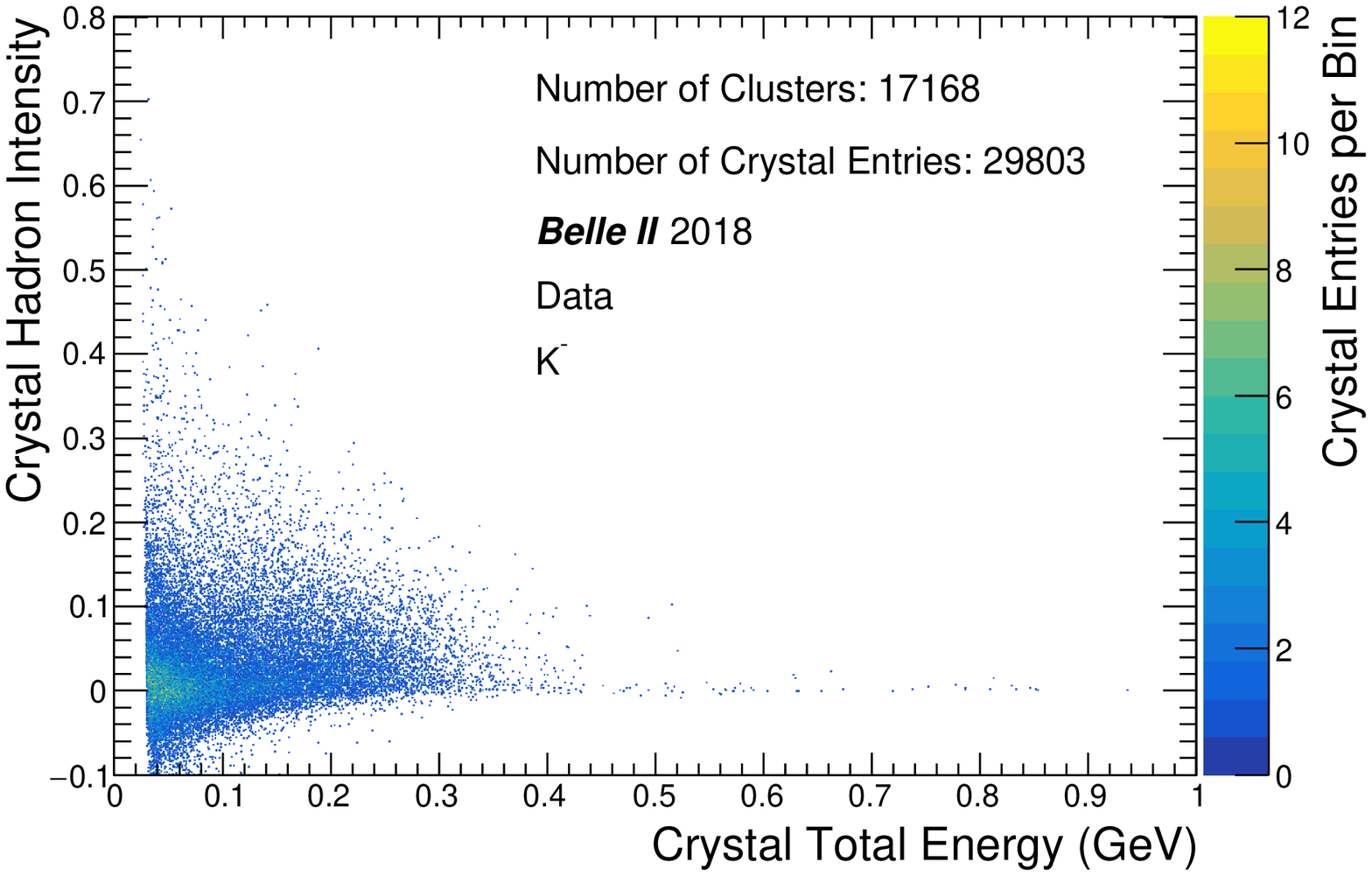}
        \caption{}
        \end{subfigure}
\begin{subfigure}[t]{.49\textwidth}
  \includegraphics[width=1\linewidth]{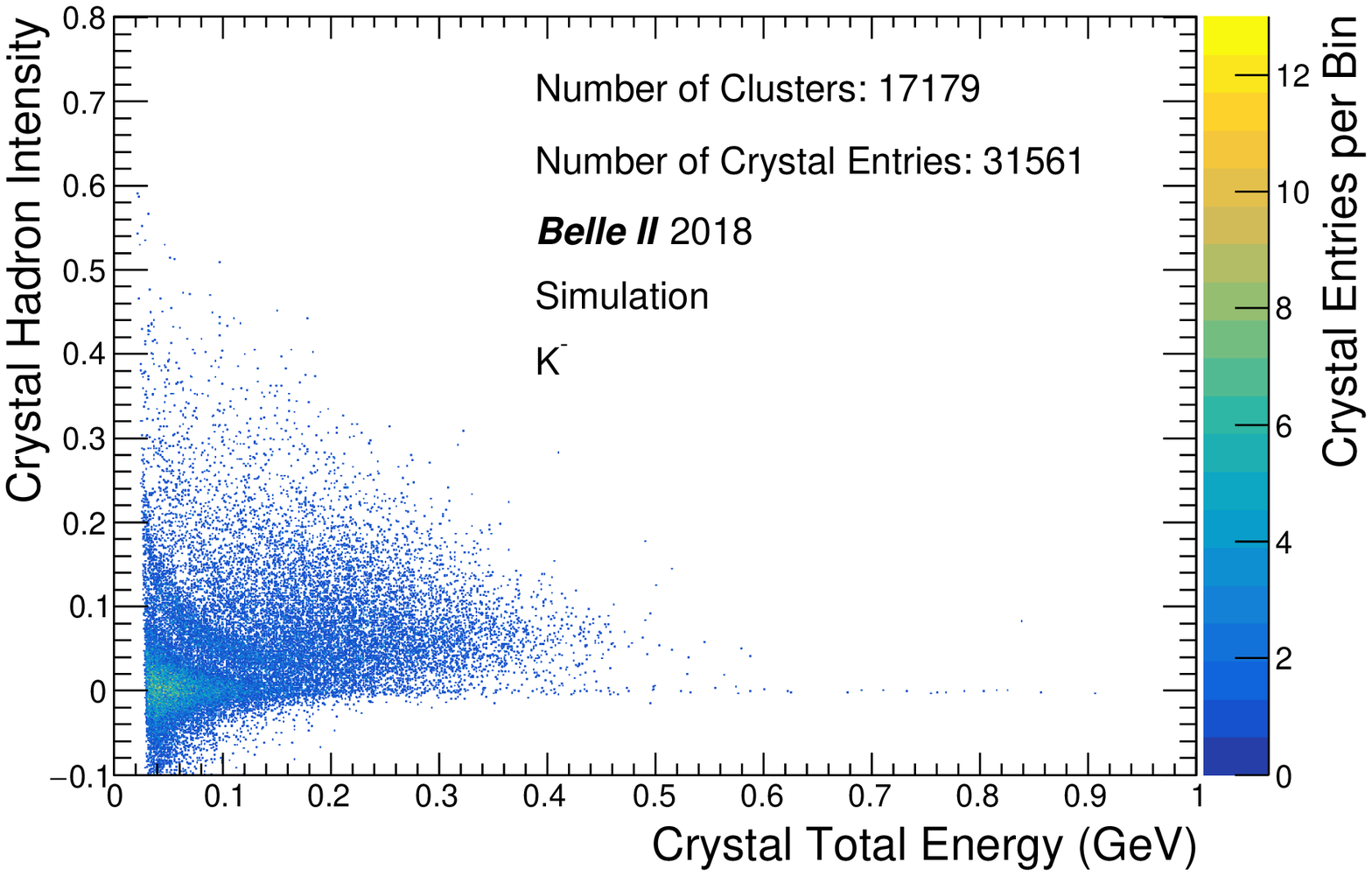}
        \caption{}
\end{subfigure}

\caption{Crystal hadron intensity vs crystal energy distributions for crystals in calorimeter clusters produced by charged kaons in the momentum range 0.3 $<$ \plab{} $\leq 0.5$ GeV/c,  a) \kp{} data b) \kp{}  simulation c) \kn{} data  d) \kn{} simulation. }
\label{f_kaonPulseShapes}
\end{figure}

This asymmetry can be visualized in Figure \ref{f_kaonPulseShapes} showing the crystal hadron intensity vs. crystal energy for crystals in calorimeter clusters produced by charged kaons in the momentum range $0.3<$ \plab{} $<0.5$ GeV/c.  Comparing the distributions for the different charges, it is observed in both data and simulation that only \kn{} results in a significant presence of crystals with large hadron scintillation emission.  The abundance of crystals with hadron pulse shapes in the \kn{} sample originate from secondary hadrons, such as protons, emitted in the decays of secondary hyperons, which are produced by the \kn{} hadronic interactions.  The charge-conjugate interactions however, are suppressed for the \kp{} resulting in very few crystals with significant hadron intensity values to be present in the \kp{} sample.  The small sample of crystals in the \kp{} distributions with hadron-like pulse shapes are found, using simulation truth, to mainly originate from hadronic interactions initiated by a secondary pion, which was emitted from the \kp{} decaying in the calorimeter.   The \kp{} vs. \kn{} asymmetry observed in the \csi{} pulse shapes was found to be maximal at low momentum.   As the kaon momenta exceeds $1$ GeV/c, the energy is sufficient for numerous \kp{} hadronic interactions to be above threshold, resulting in the pulse shape distributions to become more charge symmetric and appear similar to the charged pion distributions shown in Sections \ref{sec_chargedpion} \cite{Longo_thesis}. 
 
\subsection{Protons and anti-protons}

A proton and anti-proton control sample in the lab momentum range of 0.3-1 GeV/c was selected using \dedx{} information measured by the \belleII{} central drift chamber \cite{Longo_thesis}.  Protons and anti-protons in this momentum range generate unique \csi{} pulse shape signatures.  Proton's are unlike the particles in the previous samples studied because in this case the primary track can directly produce hadron scintillation component emission.  This is due to the significant mass of protons relative to pions and kaons resulting in the ionization \dedx{} of protons at the equivalent momenta to be significantly higher.

\begin{figure}[!ht]
\centering

\begin{subfigure}[t]{.49\textwidth}
  \includegraphics[width=1\linewidth]{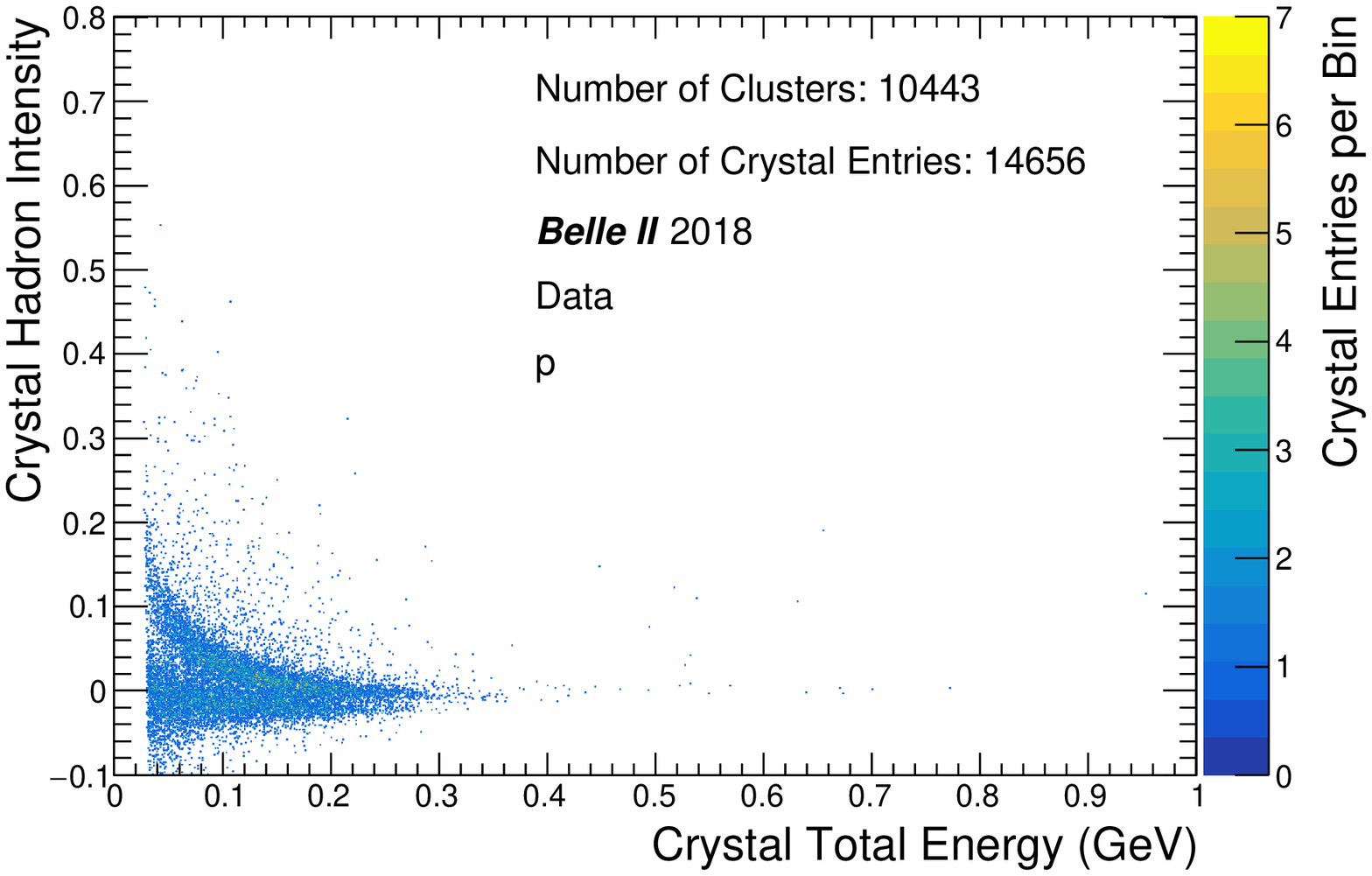}
        \caption{}
        \end{subfigure}
\begin{subfigure}[t]{.49\textwidth}
  \includegraphics[width=1\linewidth]{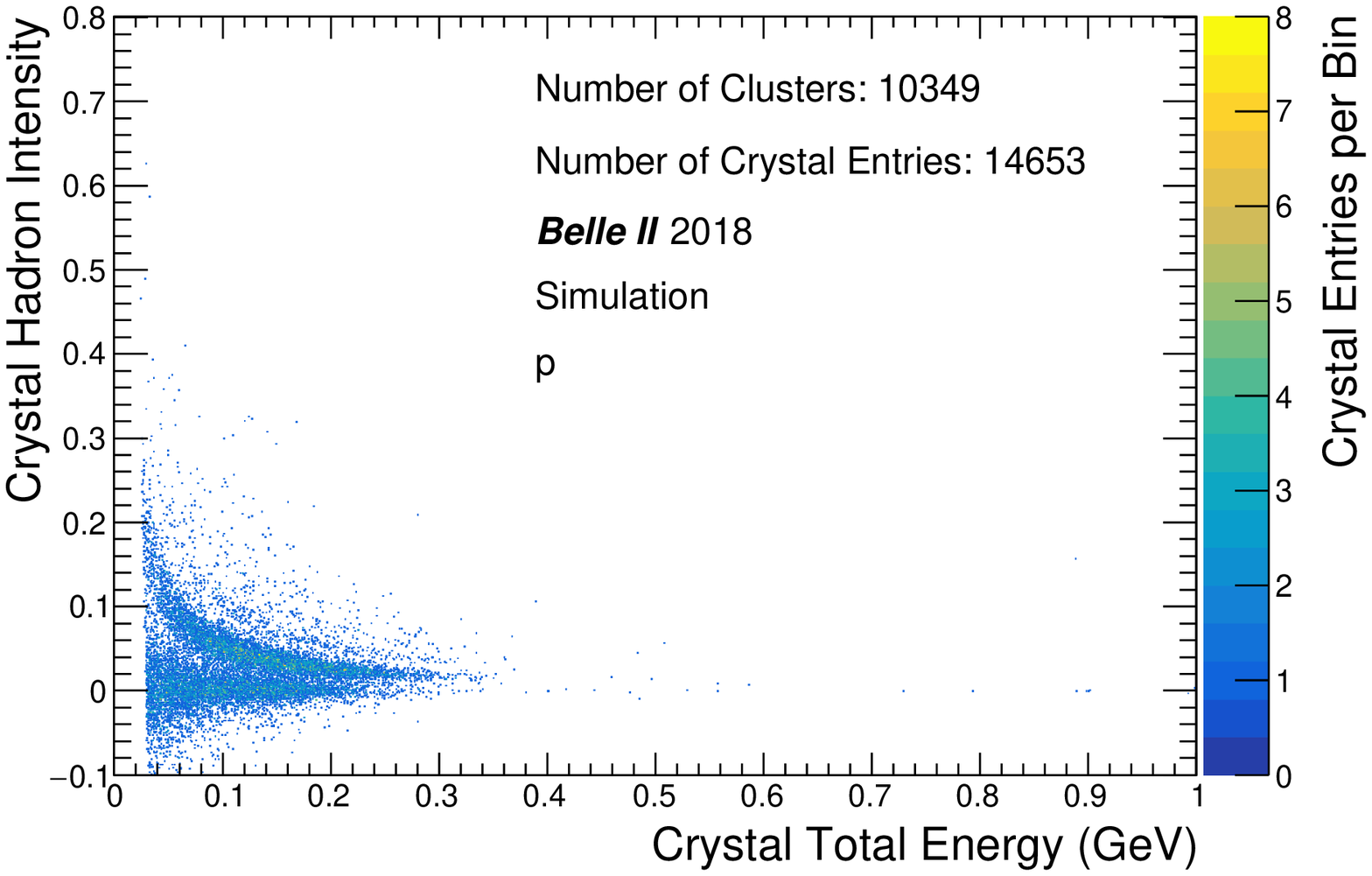}
        \caption{}
\end{subfigure}

\begin{subfigure}[t]{.49\textwidth}
  \includegraphics[width=1\linewidth]{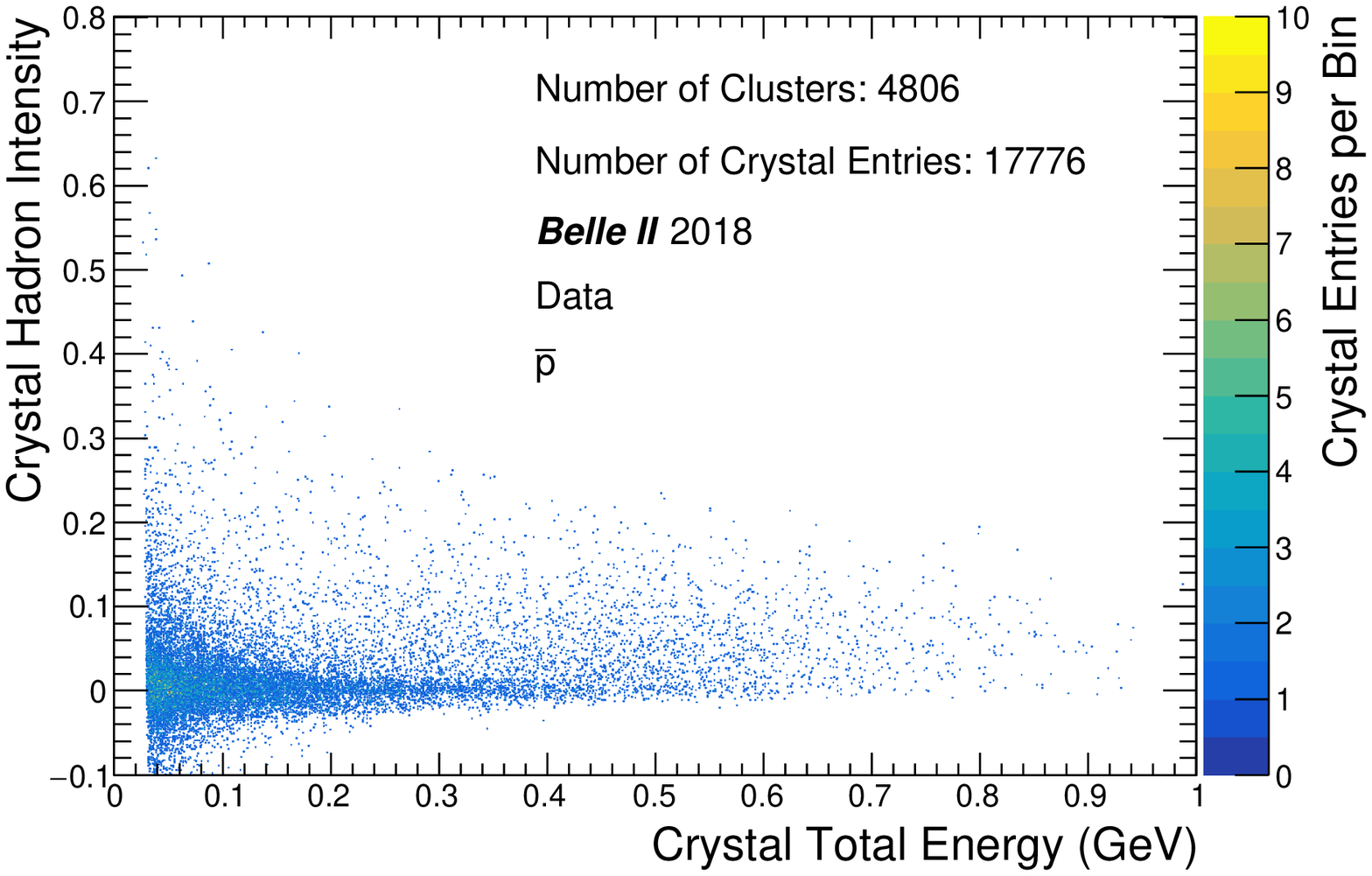}
        \caption{}
        \end{subfigure}
\begin{subfigure}[t]{.49\textwidth}
  \includegraphics[width=1\linewidth]{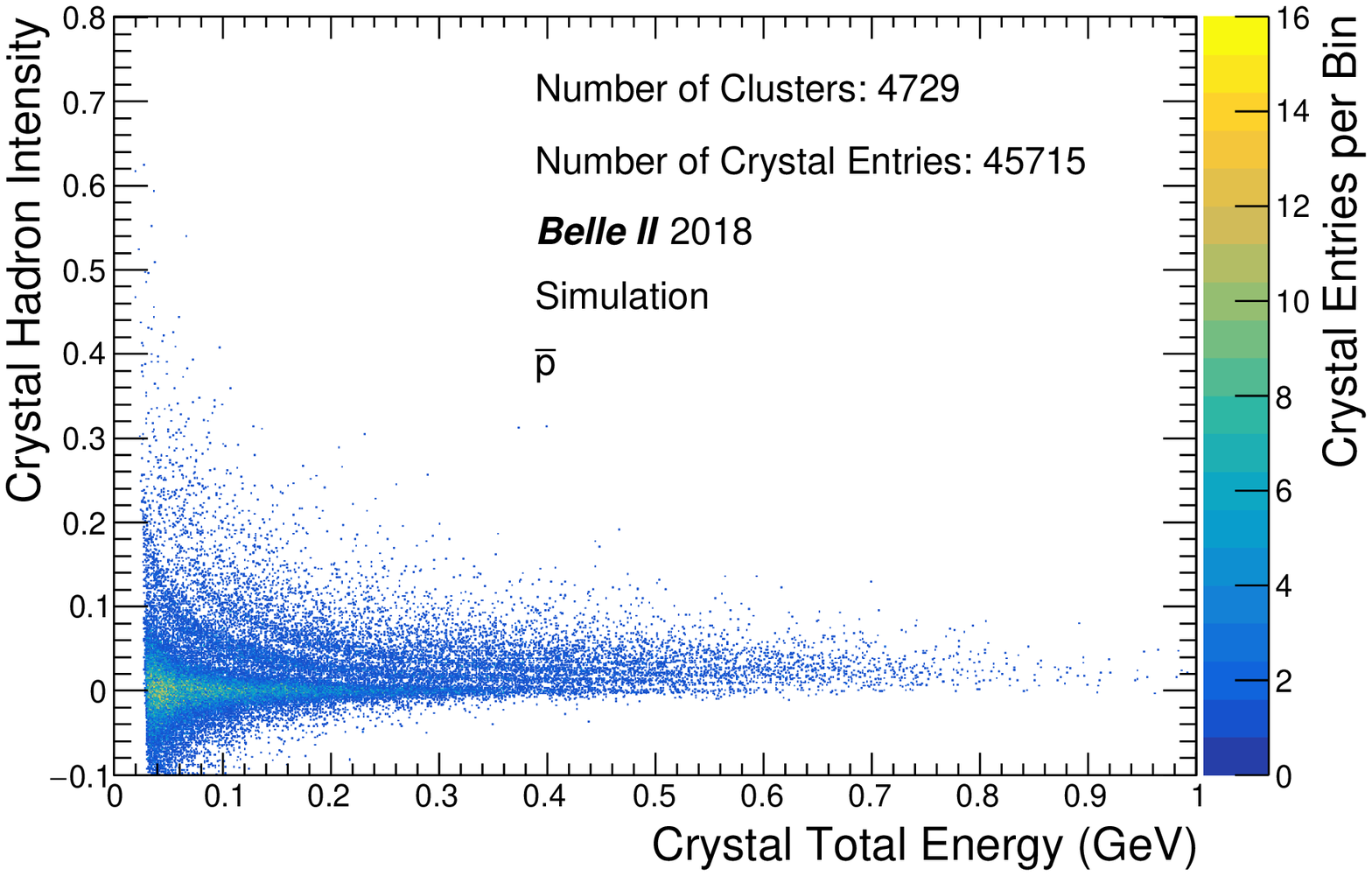}
        \caption{}
\end{subfigure}

\caption{Crystal hadron intensity vs crystal energy distributions for crystals in calorimeter clusters produced by protons in the momentum range \plab{} $0.3-1$ GeV/c, a) \pp{} data b) \pp{}  simulation c) \ap{} data  d) \ap{} simulation. }
\label{f_protonPulseShapes}
\end{figure}

In Figure \ref{f_protonPulseShapes} the crystal hadron intensity vs. crystal energy distributions are shown for crystals from calorimeter clusters produced by protons with \plab{} in the range 0.3-1 GeV/c.  In \pp{} distributions shown in Figure \ref{f_protonPulseShapes} two bands are observed both in data and simulation.  The double band structure was also displayed in the proton testbeam studies in reference \cite{Longo_2018} and arises from the possibility of the primary proton ionizing across multiple crystals.  If the \pp{} ionizes initially through a crystal at a momentum large enough such at the ionization \dedx{} of the \pp{} is not yet significant enough to produce hadron scintillation light output, then this results in the energy deposit to have a photon-like pulse shape and would correspond to the crystals in the lower band.  After escaping the initial crystal volume, the proton will then enter an adjacent crystal where it then is likely to deposit its remaining kinetic energy.   When this occurs the proton becomes highly ionizing and produces hadron light output, resulting in the energy deposit in the adjacent crystal to have a pulse shape residing on the single proton band.

 Comparing the \pp{} and \ap{} distributions, a charge asymmetry due to \ap{} annihilation is clearly observed in the pulse shape distributions.  The \ap{} distributions display similar features to the pions studied in Section \ref{sec_chargedpion}.   This is expected as anti-proton annihilation typically results in the emission of numerous charged pions \cite{Klempt2}.  The pions emitted are then likely to ionize across multiple crystal widths.  This accounts for the significant presence of crystals with energy below 0.2 GeV and photon-like pulse shapes, which are observed in the \ap{} data and simulation.  A scatter of crystals in the multi-hadron pulse shape region (crystal energy $>0.2$ GeV and hadron intensity above 0.02) are also present in the \ap{} sample.  These energy deposits originate from numerous hadrons emitted from nuclear de-excitations.  Comparing the \ap{} data and simulation, it is observed that the hadron intensity values of the crystals in the multi-hadron pulse shape region are on average lower in simulation relative to data.  In addition, the average number of crystals per cluster in simulation for \ap{} is over twice that of data.  This is interesting because for the higher momentum pions discussed in Sections \ref{sec_chargedpion}, reasonable agreement in data and simulation was observed for these quantities.  This suggests that the GEANT4 modelling of the secondary particle emission for \ap{} annihilation can be improved.  Further investigation into this discrepancy is beyond the scope of this work however, this result demonstrates how pulse shape discrimination can be used as a tool to evaluate and improve the simulation modelling of hadronic interactions.

\section{Particle Identification with Pulse Shape Discrimination}
\label{sec_PIDwPSD}

The results presented in Section \ref{sec_controlSamples} illustrate that the scintillation pulse shapes of the crystals in a calorimeter cluster are determined by the types of secondary particles produced in the cluster.  Furthermore, as the composition of secondary particles is dependent on the material interaction, and by extension, the type of primary particle, this allows the scintillation pulse shapes of the crystals in a calorimeter cluster to be used to improve particle identification.  In this section we demonstrate that through \csi{} pulse shape discrimination, significant improvements in \kl{} vs. photon identification are achieved over current techniques, which are restricted to characterization using only the spatial distribution of energy in the cluster. 

\subsection{Pulse Shape Discrimination based Multivariate Classifier}

A single calorimeter cluster can typically have multiple waveforms associated with it, one for each crystal in the cluster with energy above 30 MeV.  To condense the information contained in the multiple waveforms into a single quantity that characterizes the cluster, a multivariate classifier was trained to use the crystal level information to classify the cluster as an electromagnetic or hadronic shower.  For each crystal in the cluster that has a waveform recorded offline and where the offline fit to the waveform has a good $\chi^2$, the crystal level quantities listed below are used as classifier inputs.  

\begin{sloppy}
\begin{itemize}
\item Crystal energy computed by multi-template offline fit.
\item Crystal hadron intensity computed by multi-template offline fit.
\item Crystal offline fit type.  In addition to the Photon+Hadron template fit described Section \ref{sec_CalorDes}, fit hypotheses for a Photon+Hadron+out-of-time photon and Photon+Diode-crossing are also implemented and are attempted if the Photon+Hadron template fit results in a poor $\chi^2$.  The Photon+Hadron+out-of-time photon fit models waveforms with an additional energy deposit that is shifted in time relative to the time of the primary energy deposit and the Photon+Diode-crossing fit models the scenario where energy is directly deposited in the PIN diodes.
\item Crystal energy computed by a photon template fit done online in FPGA.
\item Location of crystal centre relative to the cluster centre.
\item Crystal weight computed by clustering algorithm to measure association of crystal to the cluster on a scale of $0-1$.
\end{itemize}
\end{sloppy}

The classifier trained is the FastBDT classifier described in reference \cite{Keck2017}.   Training was completed with particle gun samples of photons, \kl{} and anti-neutrons.  The classifier was trained such that the classifier output response to photons is 1.0 while the response to \kl{} and anti-neutrons is 0.0.

\subsection{Classifier Response to Control Samples of Photons and \kl{}}

To evaluate the classifier performance, control samples of photons and \kl{}'s were selected from \belleII{} data. The photon control sample was selected using  $e^+ e^- \rightarrow \mu^+ \mu^- (\gamma)$ events, which allowed a clean sample of photons in the momentum range $0.05 - 7$ GeV/c to be selected.  A control sample of \kl{}'s was kinematically selected using the process  $e^+ e^- \rightarrow \phi (\gamma) \rightarrow K_S^0 K_L^0 (\gamma)$.  For this selection, a \kspipi{} is required to be reconstructed in addition to two neutral calorimeter clusters.  One cluster is required to have energy above 4 GeV in the centre-of-mass frame and is identified as the initial state radiation (ISR) photon.  By applying energy conservation, the magnitude of the momentum of the \kl{} candidate is computed.  The momentum direction of the \kl{} is defined by the location of the second calorimeter cluster.  To further constrain the event to be consistent with the process $e^+ e^- \rightarrow \phi (\gamma) \rightarrow K_S^0 K_L^0 (\gamma)$, we require the invariant mass of the $K_S^0$ and $K_L^0$ to be consistent with the $\phi$ mass.  From this selection a sample of \kl{} in momenta range 2-4.5 GeV/c was selected.

In Figure \ref{f_PSDClassifier} the classifier response is shown for the selected control samples of photons and \kl{}'s with data and simulation overlaid. In Figure \ref{f_PSDClassifier_KL} the classifier response to \kl{}'s is observed to be peaking in data and simulation near 0.0, indicating that the classifier is correctly classifying the significant fraction of \kl{}'s as hadronic showers.  Conversely, in Figure \ref{f_PSDClassifier_photon} the classifier response to photons is observed to be peaking near 1.0 in data and simulation, indicating that a significant fraction of the photons are correctly classified as electromagnetic showers with significant likelihood.  The peaking nature of the distributions in Figure \ref{f_PSDClassifier} demonstrate the high degree of separation that can be achieved by applying pulse shape discrimination for hadronic vs. electromagnetic shower identification.  

In Figure \ref{f_PSDClassifier_KL} a small sample of \kl{} candidates are observed to have classifier response above 0.8.  By studying truth information of the simulation, the \kl{} in this region correspond to mis-identified \kl{} from background processes such as \eeto{}\kaon{}\kbar{}$\pi^0(\gamma)$ or \kaon{}\kbar{}$\eta$ $(\gamma)$.  In these events the \kspipi{} and the radiated photon satisfied the selection requirements and one of the photons from the decay of the the \pio{} or $\eta$ is mis-identified as a \kl{} by the kinematic selection applied. This known photon background is removed from the \kl{} selected from the ISR sample when computing the efficiency measurements completed in Section \ref{classPerf} by excluding \kl{} candidates in this sample with  classifier output above 0.8.  The bias introduced by this cut on the \kl{} efficiency is shown in Section \ref{classPerf} to be minimal relative to the statistical error of the measurement.  This is demonstrated by the measurements presented in Section \ref{classPerf} Figure \ref{f_MVAPerformance_p} as \kl{} efficiency measured for this sample is shown to be in agreement with independent \kl{} samples selected from particle gun and $B^0 \bar{B}^0$ events, which do not impose this cut.

 \begin{figure}[!ht]
\centering

\begin{subfigure}[t]{.49\textwidth}
  \includegraphics[width=1\linewidth]{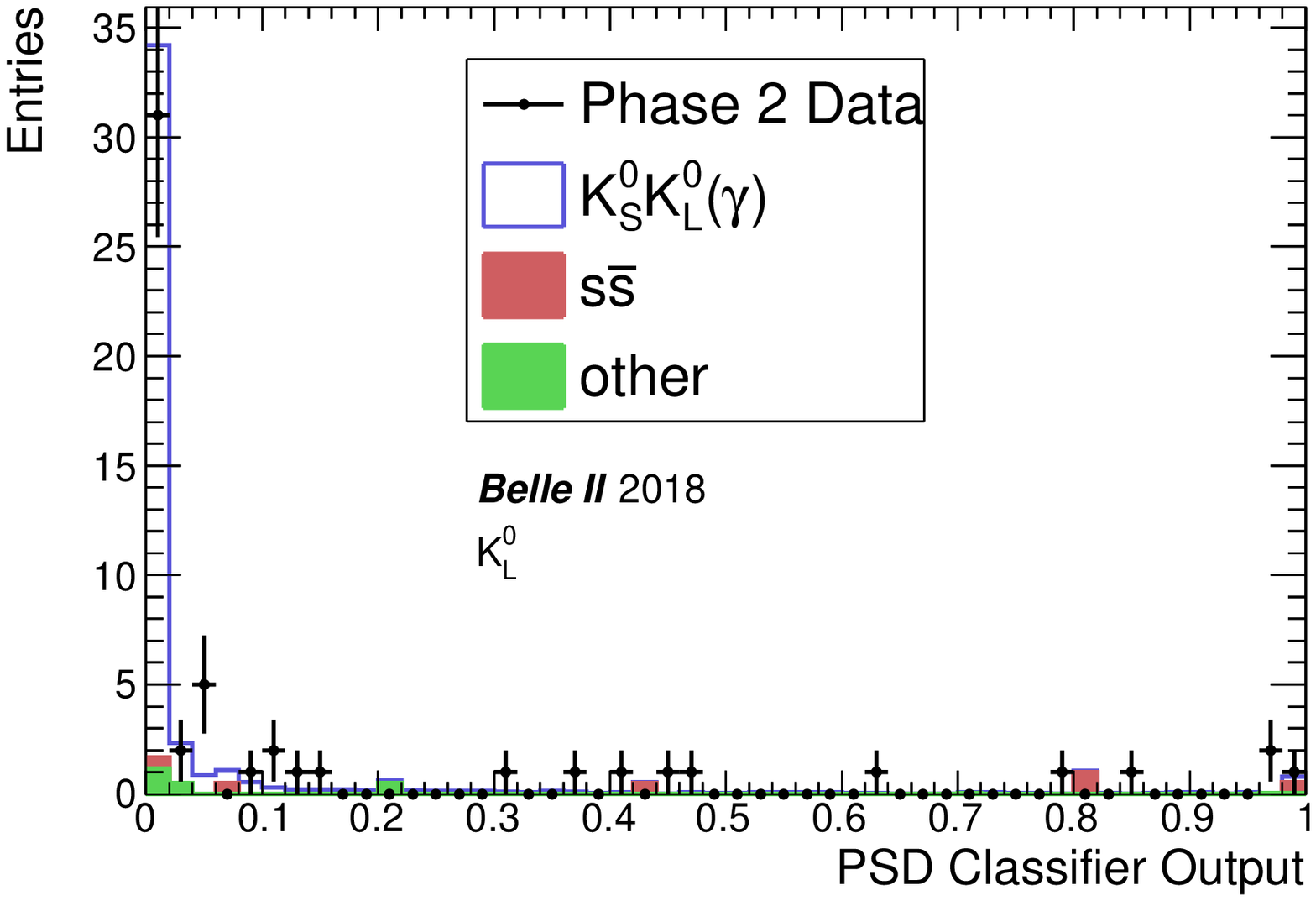}
        \caption{}
        \label{f_PSDClassifier_KL}
\end{subfigure}
\begin{subfigure}[t]{.49\textwidth}
  \includegraphics[width=1\linewidth]{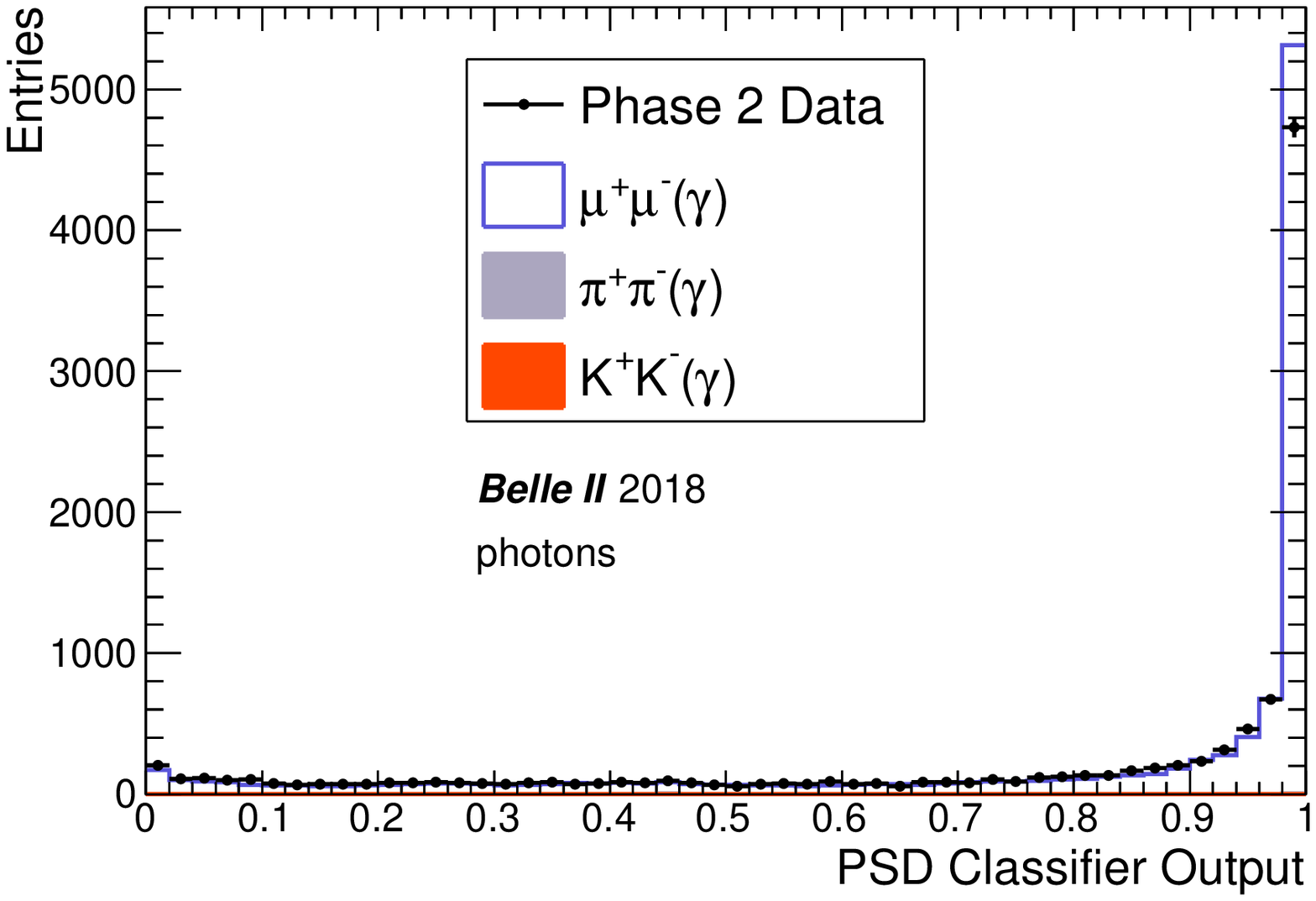}
        \caption{}
                \label{f_PSDClassifier_photon}
        \end{subfigure}

\caption{Classifier response to control samples of a) \kl{}'s and b) photon's.}
\label{f_PSDClassifier}
\end{figure}

\subsection{Impact of Pulse Shape Discrimination on Classifier Response}

To  further understand the behaviour of the classifier and confirm that the classifier response is driven by the pulse shape information, a sample of \kl{} was selected from $B^0 \bar{B}^0$ simulation.  This \kl{} selection applied truth information to select calorimeter clusters, which are produced from \kl{} emitted from a $B$ decay chain in the event.  The \kl{}'s selected in this sample also span a wide momentum range down to 0.2 GeV/c, allowing for the classifier performance in this momentum region to be measured in the following section.

For the \kl{} from $B^0 \bar{B}^0$ sample, the distribution of the classifier response was similar to the \kl{} sample shown in Figure \ref{f_PSDClassifier_KL} such that the distribution was peaking near 0.0.  With this sample however we show that the classifier response is driven by the input pulse shape information.  This is demonstrated by Figure \ref{f_KLBBbyHadInt} showing the distribution of the crystal hadron intensity vs. crystal energy for the crystals in the \kl{} from $B^0 \bar{B}^0$ clusters, divided into samples using the output of the classifier.  By scanning from Figure \ref{f_KLBBbyHadInt_hadlike} to Figure \ref{f_KLBBbyHadInt_photlike} the types of pulse shapes that are present in \kl{} clusters classified as hadron-like to photon-like can be visualized. 

Comparing Figure \ref{f_KLBBbyHadInt_hadlike} to Figure \ref{f_KLBBbyHadInt_photlike}, it is observed that clusters classified as less hadron-like correspond to clusters that contain crystals with pulse shapes that are less hadron-like.  From Figure \ref{f_KLBBbyHadInt_hadlike} it is shown that if the cluster contains a crystal with a pulse shape in the multi-hadron region, then the cluster will be classified as hadronic with significant likelihood, contributing to the peaking structure near 0.0 in Figure \ref{f_PSDClassifier_KL}.   This is in contrast to \kl{} clusters mis-classified as photon-like.  As shown by Figure \ref{f_KLBBbyHadInt_photlike}, the \kl{} clusters classified as photon-like are observed to contain only crystals with photon-like pulse shapes.  This demonstrates that these clusters are mis-classified because the pulse shapes of the crystals in the clusters are photon-like and thus in terms of pulse shape discrimination, the cluster appears as an electromagnetic shower.  

\begin{figure}[!ht]
\centering

\begin{subfigure}[t]{.49\textwidth}
  \includegraphics[width=1\linewidth]{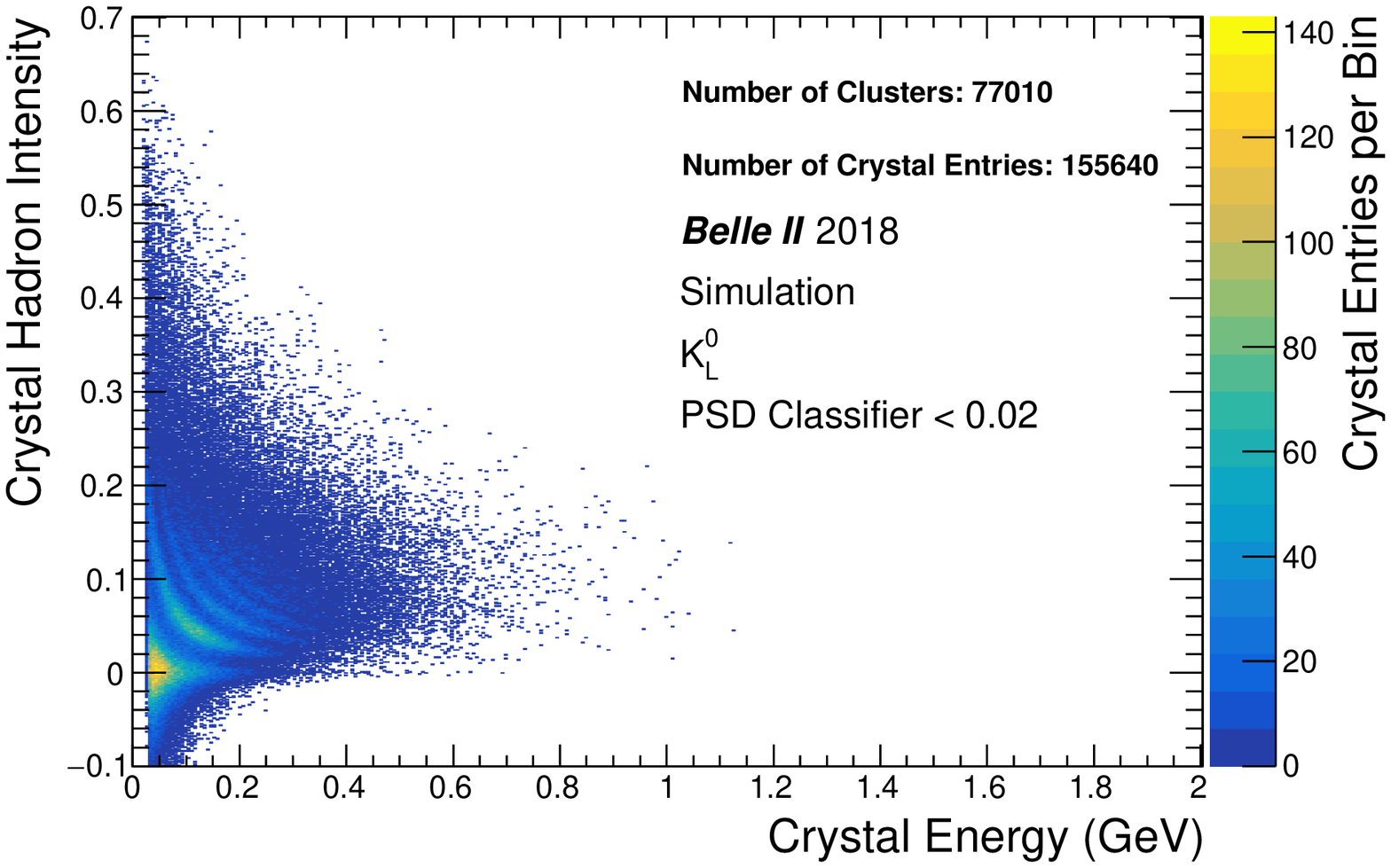}
        \caption{}
                \label{f_KLBBbyHadInt_hadlike}
        \end{subfigure}
\begin{subfigure}[t]{.49\textwidth}
  \includegraphics[width=1\linewidth]{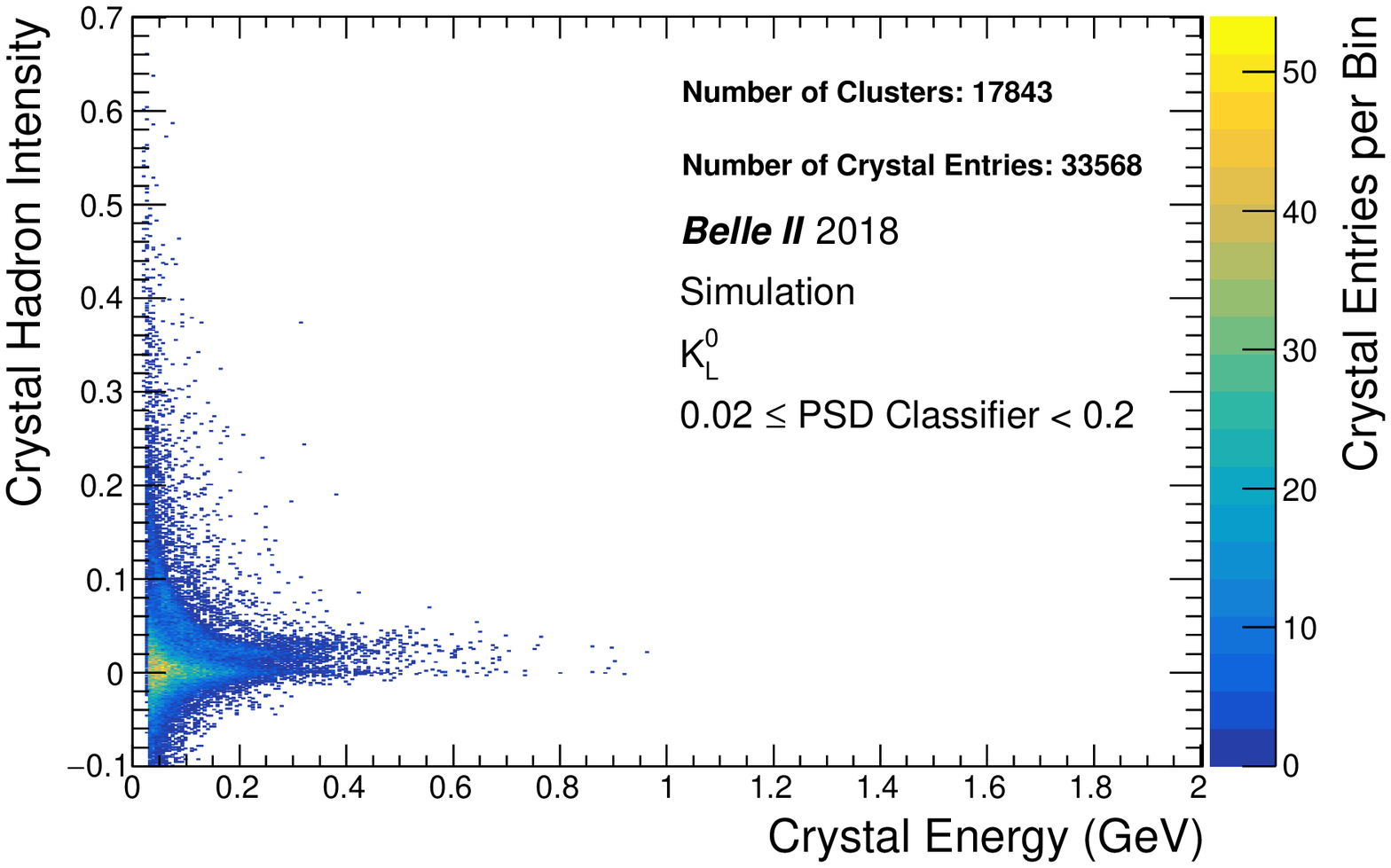}
        \caption{}
\end{subfigure}

\begin{subfigure}[t]{.49\textwidth}
  \includegraphics[width=1\linewidth]{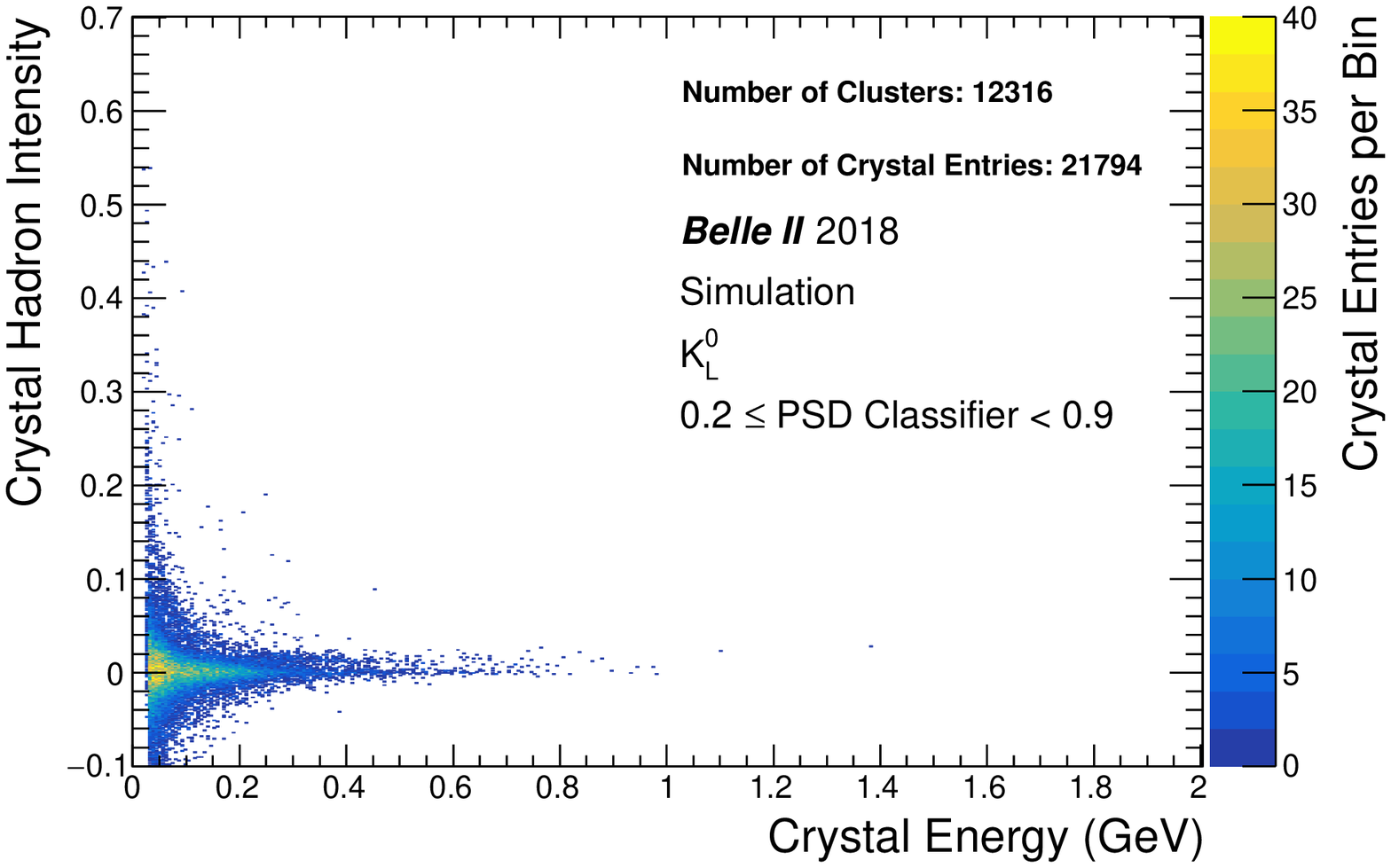}
        \caption{}
        \end{subfigure}
\begin{subfigure}[t]{.49\textwidth}
  \includegraphics[width=1\linewidth]{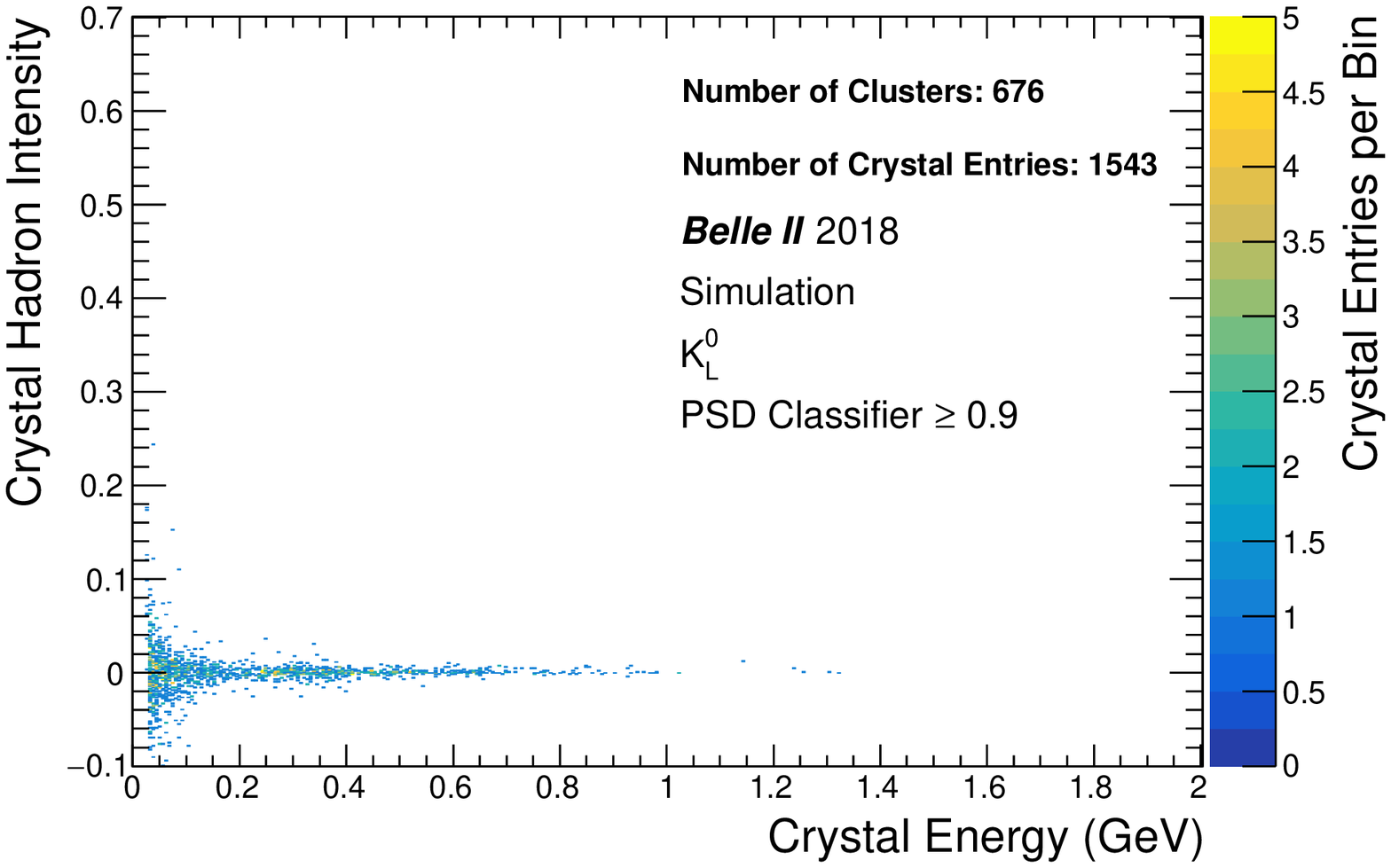}
        \caption{}
\label{f_KLBBbyHadInt_photlike}
\end{subfigure}

\caption{Crystal hadron intensity vs. crystal energy distribution for crystals in clusters from \kl{} from the $B^0 \bar{B}^0$ sample, divided into four samples of classifier response.}
\label{f_KLBBbyHadInt}
\end{figure}

\subsection{Classifier Performance}
\label{classPerf}

To present the performance of the classifier a threshold of $<0.1$ is imposed on the PSD Classifier Output to classify a cluster as a hadronic shower.  The \kl{} efficiency and photon-as-hadron fake-rate are then measured as a function of cluster energy and particle momentum.   This approach is advantageous because, although for photons the cluster energy is equal to the photon momentum, for \kl{} the cluster energy is only loosely correlated with the \kl{} momentum due to energy leakage from the invisible component of the hadronic shower.  In addition the relative components of the \kl{} interaction cross section will vary as a function of the \kl{} momentum and thus  the classifier performance should be studied as a function of the cluster energy and particle momentum.

\begin{figure}[!ht]
\centering
  \includegraphics[width=0.75\linewidth]{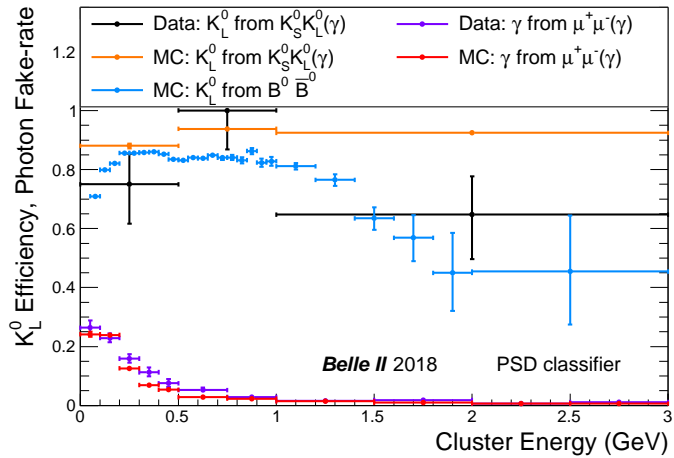}
\caption{Measurement of the \kl{} efficiency and photon-as-hadron fake-rate for the PSD classifier as a function of cluster energy for several control samples of \kl{}, and photons.  Errors bars correspond to statistical errors.}
        \label{f_MVAPerformance_E}
\end{figure}

Figure \ref{f_MVAPerformance_E} presents measurements of the \kl{} efficiency and photon-as-hadron fake-rate as a function of cluster energy for the selected control samples of photons and \kl{}'s.  Focusing on the photon-as-hadron fake-rate shown in Figure \ref{f_MVAPerformance_E}, a dependence on the cluster energy is observed.  Above 1 GeV the photon-as-hadron fake-rate is measured to be below 2\% in data and simulation.  As the photon momentum drops below 0.5 GeV/c the photon-as-hadron fake rate increases in data and simulation to a maximum of 25\% at 0.1 GeV/c.  This energy dependence is driven by two factors.  The first is due to the 30 MeV waveform readout threshold causing lower energy photons to have less crystals with pulse shape information available.  The second factor is due to the degradation of the hadron intensity resolution at lower crystal energies, which arises from the decrease in the signal-to-noise ratio of the waveforms.  This effect was observed previously in Section \ref{photoncrystals} Figure \ref{f_photonPulseShapes} by the broadening of the hadron intensity values with decreasing crystal energy.  This degradation in resolution increases the difficulty to definitively classify a crystal energy deposit as hadronic or electromagnetic. 

In Figure \ref{f_MVAPerformance_E} multiple \kl{} efficiency measurements are overlaid.  These measurements are complementary as together they span the full \kl{} momentum range of interest for the \belleII{} experiment.  Beginning by studying the \kl{} from ISR sample, it is observed that at cluster energies below 1 GeV the \kl{} efficiency is measured to be above 80\% in data and simulation.  At cluster energies above 1 GeV, the \kl{} efficiency in data is observed to be $2 \sigma$ below the value in simulation. Note the error bars in Figure \ref{f_MVAPerformance}  correspond to statistical errors.  With a larger data sample the significance of this difference in data and simulation can be verified.  If confirmed, a potential source of this discrepancy could be from the modelling of the \kl{} hadronic interactions in \csi{} by GEANT4.  If the simulated cross section for \kl{} interactions that produce final states with $\pi^0$'s is over estimated then this could result in such a discrepancy.  This is because in these interactions the full \pio{} energy is typically absorbed in the form of an electromagnetic shower and thus if a \pio{} is produced the resulting cluster will likely be on the higher end of the cluster energy spectrum.  In addition the significant electromagnetic shower component of the hadronic shower would result in the cluster to appear more photon-like in to the pulse shape classifier.  This effect was verified through simulation truth information to be the cause of the drop in efficiency at larger clusters energies, which is observed in Figure \ref{f_MVAPerformance_E} for the \kl{} from $B^0 \bar{B}^0$ sample.

\subsection{Comparison to a Shower-shape based Approach}

To demonstrate the improvement in particle identification that can be achieved with pulse shape discrimination, this section compares the PSD classifier performance to two shower-shape approaches to \kl{} vs photon identification.  

The shower-shape variable \eoen{} is defined as the ratio of the energy in the centre cluster crystal to the total energy of the $3 \times 3$ crystal array centred about the central crystal.  Due to the symmetric and centrally peaking nature of electromagnetic showers, photons tend to have \eoen{} values larger than hadronic showers, where the energy in the cluster is more widely distributed.  We note that although \eoen{} is one of several shower-shapes and likely alone does not exploit the full shower-shape potential, it has been selected for use in independent optimization studies for photon selection at \belleII{} for \pio{} reconstruction \cite{SebastianThesis}.  This allows it to provide a reference to the performance of the typical shower-shape variables currently applied.

A second shower-shape approach that we compare to is an independently trained shower-shape based classifier referred to as the Zernike classifier.  The Zernike classifier is a Boosted Decision Tree trained to use the spatial distribution of the cluster energy to perform photon vs. \kl{} identification.  The inputs to the Zernike classifier are the first eleven Zernike moments of the cluster computed from the energy of the cluster crystals, in addition to the first eleven Zernike moments of the larger connected region of clusters that the local cluster might belong to.  Each of these moments corresponds to a cluster energy centroid-like quantity computed by applying the Zernike polynomials \cite{ZERNIKE1934689} as weights.  Unlike the PSD classifier and the variable \eoen{}, the input information to the Zernike classifier extends beyond the primary cluster as the second eleven inputs to the Zernike classifier incorporate information from all clusters in close spatial proximity to the main cluster.  For \kl{} identification, this boosts performance from the use of split-off information, however, this in turn limits performance of the Zernike classifier in  situations where the photon is not well isolated.

To compare the performance of the PSD classifier to the shower-shape approaches, Figure \ref{f_MVAPerformance} shows the \kl{} efficiency and corresponding photon-as-hadron fake-rate measured as a function of particle momentum for the PSD classifier (Figure \ref{f_MVAPerformance_p}), the \eoen{} variable (Figure \ref{f_MVAPerformance_E1E9}) and for the Zernike classifier (Figure \ref{f_MVAPerformance_zer}) at threshold's which set the corresponding \kl{} efficiency above 3 GeV/c to be equal to the PSD classifier efficiency.  For the measurements in Figure \ref{f_MVAPerformance}, the Zernike classifier and \eoen{} thresholds are set such that the same \kl{} efficiency as the PSD classifier is achieved.  By setting the \kl{} efficiency equal between the three methods, the performance of the methods is compared using the corresponding photon-as-hadron fake-rate, where a lower  photon-as-hadron fake-rate indicates better performance.  In Figure \ref{f_MVAPerformance} the momentum of the \kl{} from $e^+ e^- \rightarrow K_S^0 K_L^0 (\gamma)$ sample is computed using the measured $K_S^0$ and $\gamma$ momentum, and applying energy conservation. The momentum of the \kl{} from the $B^0 \bar{B}^0$ sample and the particle gun sample corresponds to the generated momentum of the simulated \kl{}.

\begin{figure}[!ht]
\centering

\begin{subfigure}[t]{.49\textwidth}
  \includegraphics[width=1\linewidth]{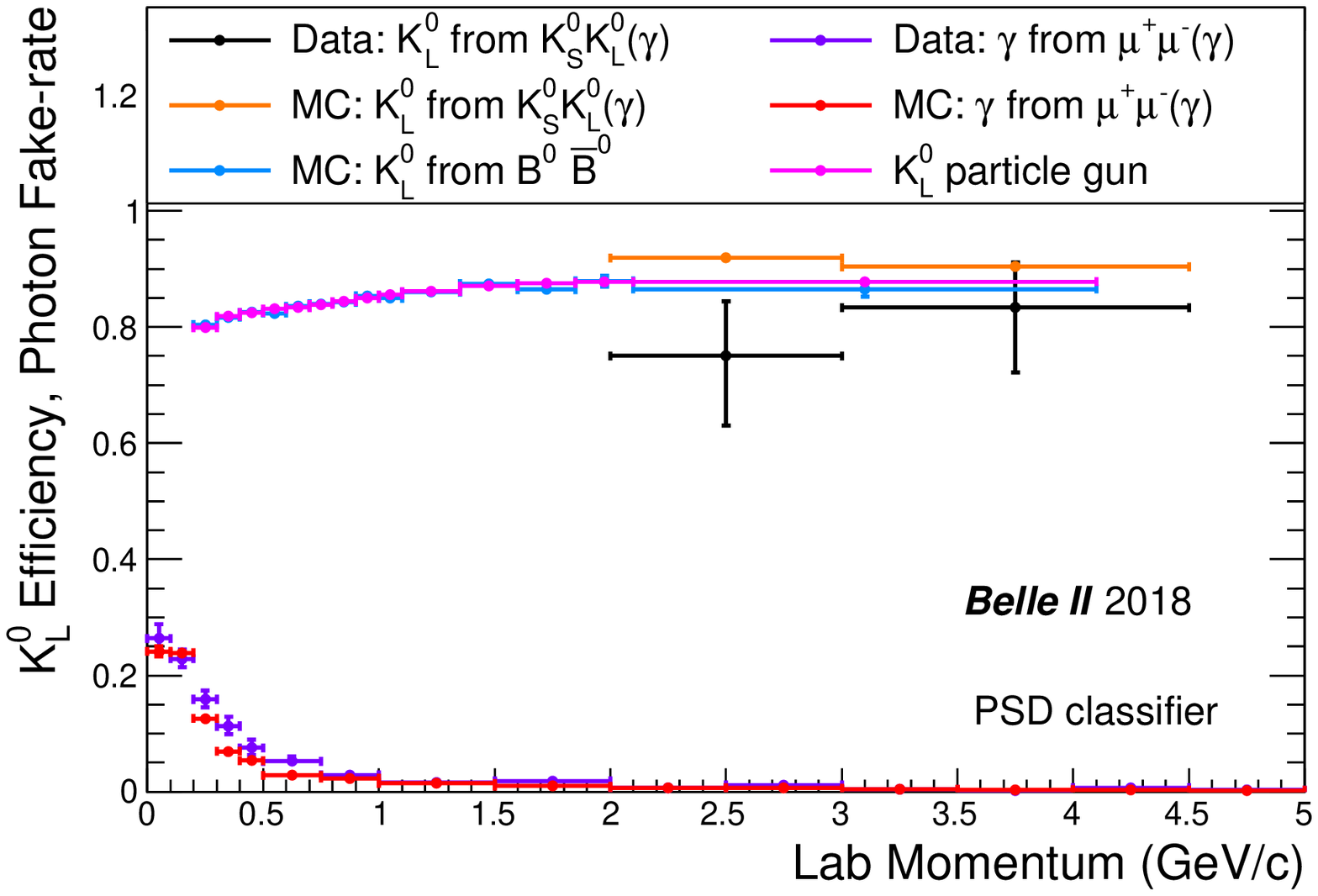}
        \caption{}
                \label{f_MVAPerformance_p}
\end{subfigure}
\begin{subfigure}[t]{.49\textwidth}
  \includegraphics[width=1\linewidth]{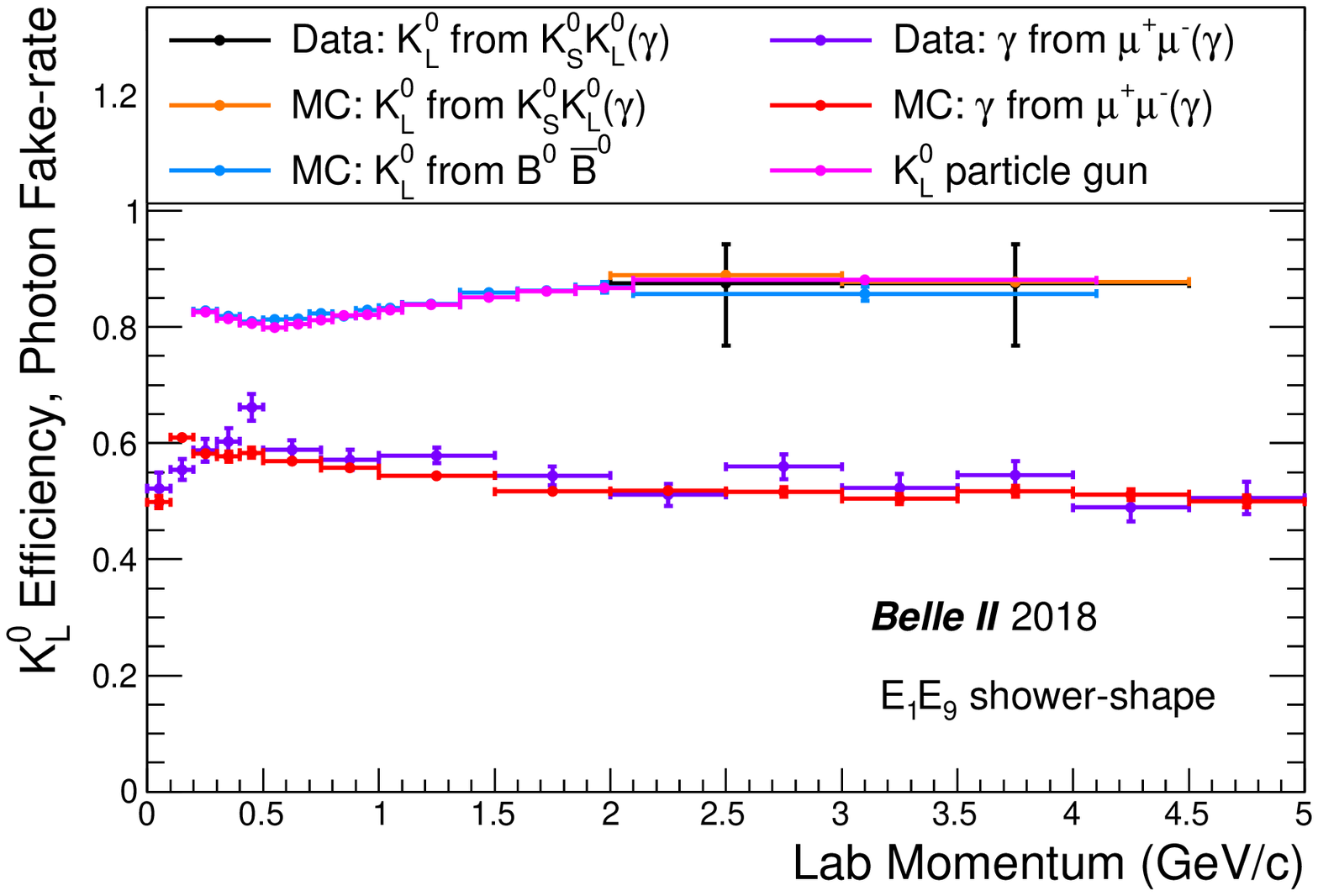}
        \caption{}
\label{f_MVAPerformance_E1E9}
        \end{subfigure}

\begin{subfigure}[t]{.49\textwidth}
  \includegraphics[width=1\linewidth]{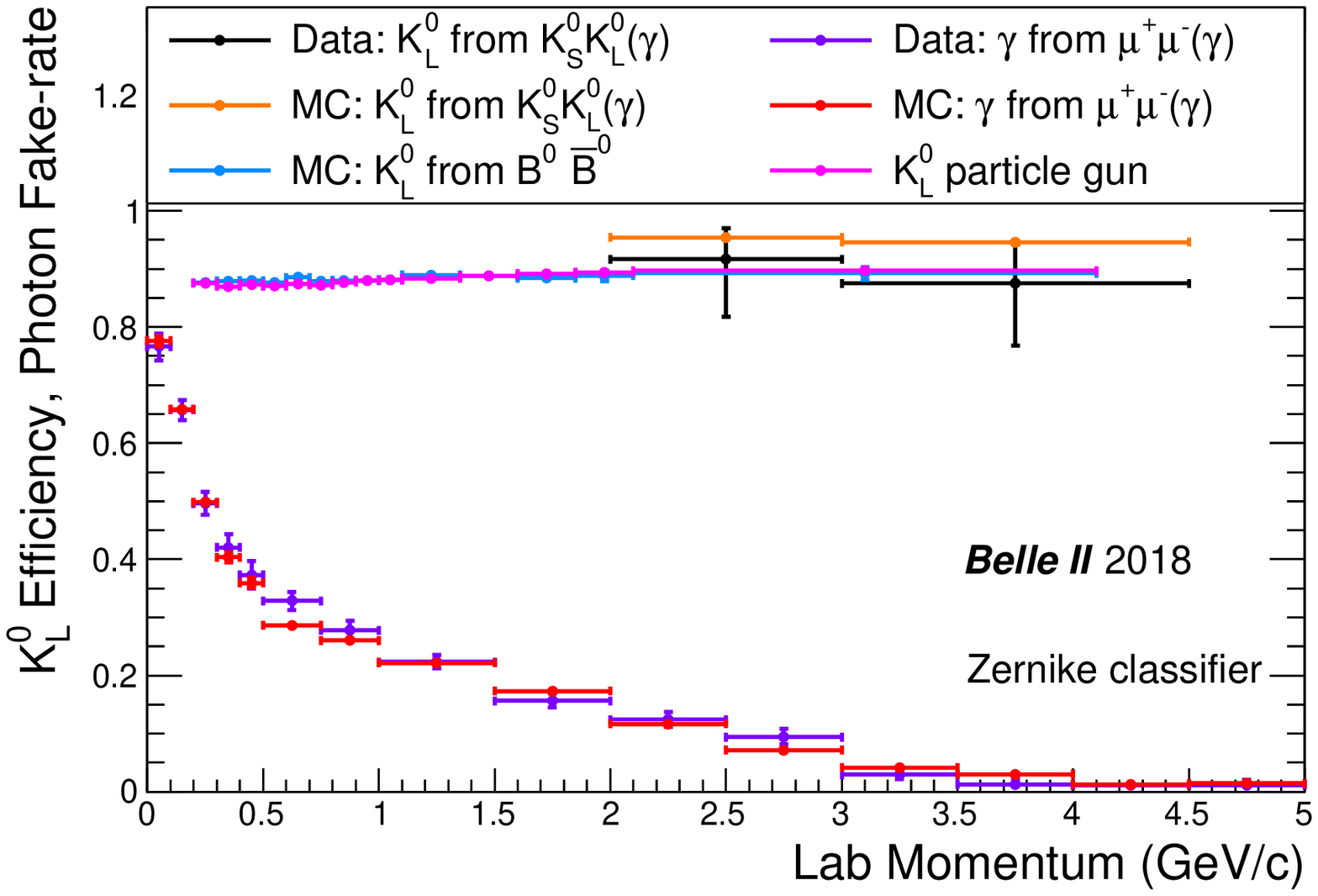}
        \caption{}
\label{f_MVAPerformance_zer}
        \end{subfigure}

\caption{Measurement of the \kl{} efficiency and photon-as-hadron fake-rate for the a) PSD classifier b) \eoen{} shower-shape c) Zernike classifier as a function of particle momentum for several control samples of \kl{}, and photon's. The Zernike classifier is an independent classifier, which uses only shower shape information.  The threshold for the \eoen{} shower-shape and Zernike classifier is set such that the same \kl{} efficiency as the PSD classifier is achieved; this allows the comparison of the performance of the classifiers to be made using the photon-as-hadron fake-rate.  Errors bars correspond to statistical errors.}
\label{f_MVAPerformance}
\end{figure}

In Figure \ref{f_MVAPerformance_E1E9} the photon-as-hadron fake-rate for the \eoen{} shower-shape variable is observed to be approximately constant as a function of the photon energy with a value of about 55\%.  Compared to the PSD classifier, which at 1 GeV/c achieves a photon-as-hadron fake-rate below 2\% and a photon-as-hadron fake-rate of 15\% at 0.2 GeV/c, the significant improvement in \kl{} vs. photon separation that can be achieved by applying \csi{} pulse shape discrimination is demonstrated.  

In Figure \ref{f_MVAPerformance_zer} the Zernike classifier is shown have a photon-as-hadron fake-rate of $23\%$ at 1 GeV/c, increasing to a fake-rate of 50\% at 0.2 GeV/c.  The improvement of the Zernike classifier over the \eoen{} shower-shape variable demonstrates that incorporating information beyond the primary cluster can improve the \kl{} vs. photon identification performance.  The PSD classifier is observed to achieved improved performance over the Zernike classifier, reducing the photo-as-hadron fake-rate by a factor of 3 at lower photon energies and a factor of 10 at photon energies above 1 GeV.  Recall that the Zernike classifier uses information not just from the primary cluster but also from any clusters in close proximity to the primary cluster in order to improve the \kl{} efficiency.    Further investigating the Zernike classifier behaviour in the photon sample we find that, even in the low multiplicity $e^+ e^- \rightarrow \mu^+ \mu^- (\gamma)$ events, a limiting factor of the photon-as-hadron fake-rate for the Zernike classifier is due to the photon isolation from the muons in the event.  Due to the PSD classifier only using information from the primary cluster, and that PSD is directly sensitive to the secondary particle composition of the cluster whereas shower-shape approaches can only have indirect inference, PSD approaches are much less sensitive to cluster isolation effects.  This also indicates an area for future studies can be to investigate the application of PSD techniques in specialized cases where shower-shape approaches are known to be limited, such as clusters from merged photons resulting in the shape of the electromagnetic shower to appear hadron-like.

\section{Summary and Conclusions}
\label{conclustions}

This paper presents the first application of \csi{} pulse shape discrimination at an electron-positron collider.  The results demonstrate that \csi{} pulse shape discrimination provides a new and effective method for improving \kl{} vs. photon separation at the \belleII{} experiment.   

The upgrade of the \belleII{} calorimeter readout with waveform digitization having 18-bit precision and a 1.7669 MHz sampling frequency enabled the implementation of pulse shape discrimination.  A multi-template fit is applied to characterize the waveform shape by measuring the fraction of scintillation emission emitted by the hadron scintillation component.  Distributions of the scintillation pulse shape of crystals from calorimeter clusters produced by control samples of $\gamma$, $\mu^+$, $\pi^\pm$, $K^\pm$ and $p/\bar{p}$ illustrated how the scintillation pulse shape can identify the secondary particle composition of the calorimeter cluster.  Comparisons between data and simulation provided further validation of the \csi{} scintillation response simulation techniques developed in reference \cite{Longo_2018} and now are applied at \belleII{}.   These studies demonstrate the potential application of pulse shape discrimination as a new tool to identify areas of improving the simulation modelling of hadronic interactions in materials.

The \kl{} efficiency and photon-as-hadron fake-rate of a multivariate classifier trained to use pulse shape discrimination to classify calorimeter clusters as hadronic or electromagnetic showers was evaluated using control samples of photons and \kl{}'s selected from \belleII{} commissioning data.  Pulse shape discrimination was shown to be responsible for the classifier performance, which was measured to achieve a \kl{} efficiency above 80\% with a corresponding photon-as-hadron fake-rate of 15\% at cluster energies of 0.2 GeV and below 2\% at cluster energies of 1 GeV.   This corresponds to a significant improvement in photon vs. \kl{} identification over a standard shower-shape approach, which for comparison was measured to have a photon-as-hadron fake-rate of 55\% in order to achieve the same \kl{} efficiency.  Comparisons to a shower-shape based classifier also demonstrated that PSD can reduce the photon-as-hadron fake-rate by a factor of 3 at photon energies of 0.2 GeV and a factor 10 at photon energies of 1 GeV.

To improve performance there are several areas for potential upgrades in the future.   A limitation of the present performance was shown to be due to the 30 MeV waveform readout threshold restricting the amount of pulse shape information available for lower energy clusters.  Implementation of waveform shape characterization at the FPGA level could remove this limitation in the future.  In addition, although this study focused on \kl{} vs. photon identification, the results presented show clear potential for the application of pulse shape discrimination to improve charged particle identification, as well as, motivating studies to explore the potential for neutron vs. \kl{} separation using pulse shape discrimination.

The improvements in the areas of neutral particle identification achieved with the application of \csi{} pulse shape discrimination at \belleII{} encourages other calorimeters constructed from pulse shape discrimination capable scintillators such as NaI(Tl) \cite{Bartle},
CsI(Na)\cite{PSD_CsIBaF}, 
pure CsI \cite{PSD_CsIpure},
PbWO$_4$ (doped) \cite{PSD_PbWO}, 
BaF$_2$\cite{PSD_CsIBaF} to consider implementing this technique.  In addition calorimeter designs for future experiments should consider extending design criteria to include pulse shape discrimination.

\section*{Acknowledgements}

\begin{sloppypar}
We thank the SuperKEKB group for the excellent operation of the
accelerator; the KEK cryogenics group for the efficient
operation of the solenoid; and the KEK computer group for
on-site computing support.
This work was supported by the following funding sources:
Science Committee of the Republic of Armenia Grant No. 18T-1C180;
Australian Research Council and research grant Nos.
DP180102629, 
DP170102389, 
DP170102204, 
DP150103061, 
FT130100303, 
and
FT130100018; 
Austrian Federal Ministry of Education, Science and Research, and
Austrian Science Fund No. P 31361-N36; 
Natural Sciences and Engineering Research Council of Canada, Compute Canada and CANARIE;
Chinese Academy of Sciences and research grant No. QYZDJ-SSW-SLH011,
National Natural Science Foundation of China and research grant Nos.
11521505,
11575017,
11675166,
11761141009,
11705209,
and
11975076,
LiaoNing Revitalization Talents Program under contract No. XLYC1807135,
Shanghai Municipal Science and Technology Committee under contract No. 19ZR1403000,
Shanghai Pujiang Program under Grant No. 18PJ1401000,
and the CAS Center for Excellence in Particle Physics (CCEPP);
the Ministry of Education, Youth and Sports of the Czech Republic under Contract No.~LTT17020 and 
Charles University grants SVV 260448 and GAUK 404316;
European Research Council, 7th Framework PIEF-GA-2013-622527, 
Horizon 2020 Marie Sklodowska-Curie grant agreement No. 700525 `NIOBE,' 
and
Horizon 2020 Marie Sklodowska-Curie RISE project JENNIFER2 grant agreement No. 822070 (European grants);
L'Institut National de Physique Nucl\'{e}aire et de Physique des Particules (IN2P3) du CNRS (France);
BMBF, DFG, HGF, MPG, AvH Foundation, and Deutsche Forschungsgemeinschaft (DFG) under Germany's Excellence Strategy -- EXC2121 ``Quantum Universe''' -- 390833306 (Germany);
Department of Atomic Energy and Department of Science and Technology (India);
Israel Science Foundation grant No. 2476/17
and
United States-Israel Binational Science Foundation grant No. 2016113;
Istituto Nazionale di Fisica Nucleare and the research grants BELLE2;
Japan Society for the Promotion of Science,  Grant-in-Aid for Scientific Research grant Nos.
16H03968, 
16H03993, 
16H06492,
16K05323, 
17H01133, 
17H05405, 
18K03621, 
18H03710, 
18H05226,
19H00682, 
26220706,
and
26400255,
the National Institute of Informatics, and Science Information NETwork 5 (SINET5), 
and
the Ministry of Education, Culture, Sports, Science, and Technology (MEXT) of Japan;  
National Research Foundation (NRF) of Korea Grant Nos.
2016R1\-D1A1B\-01010135,
2016R1\-D1A1B\-02012900,
2018R1\-A2B\-3003643,
2018R1\-A6A1A\-06024970,
2018R1\-D1A1B\-07047294,
2019K1\-A3A7A\-09033840,
and
2019R1\-I1A3A\-01058933,
Radiation Science Research Institute,
Foreign Large-size Research Facility Application Supporting project,
the Global Science Experimental Data Hub Center of the Korea Institute of Science and Technology Information
and
KREONET/GLORIAD;
Universiti Malaya RU grant, Akademi Sains Malaysia and Ministry of Education Malaysia;
Frontiers of Science Program contracts
FOINS-296,
CB-221329,
CB-236394,
CB-254409,
and
CB-180023, and SEP-CINVESTAV research grant 237 (Mexico);
the Polish Ministry of Science and Higher Education and the National Science Center;
the Ministry of Science and Higher Education of the Russian Federation,
Agreement 14.W03.31.0026;
University of Tabuk research grants
S-1440-0321, S-0256-1438, and S-0280-1439 (Saudi Arabia);
Slovenian Research Agency and research grant Nos.
J1-9124
and
P1-0135; 
Agencia Estatal de Investigacion, Spain grant Nos.
FPA2014-55613-P
and
FPA2017-84445-P,
and
CIDEGENT/2018/020 of Generalitat Valenciana;
Ministry of Science and Technology and research grant Nos.
MOST106-2112-M-002-005-MY3
and
MOST107-2119-M-002-035-MY3, 
and the Ministry of Education (Taiwan);
Thailand Center of Excellence in Physics;
TUBITAK ULAKBIM (Turkey);
Ministry of Education and Science of Ukraine;
the US National Science Foundation and research grant Nos.
PHY-1807007 
and
PHY-1913789, 
and the US Department of Energy and research grant Nos.
DE-AC06-76RLO1830, 
DE-SC0007983, 
DE-SC0009824, 
DE-SC0009973, 
DE-SC0010073, 
DE-SC0010118, 
DE-SC0010504, 
DE-SC0011784, 
DE-SC0012704; 
and
the National Foundation for Science and Technology Development (NAFOSTED) 
of Vietnam under contract No 103.99-2018.45.
\end{sloppypar}

\bibliography{PSD-BelleIICalorGroupPaper_bib}

\begin{thebibliography}{10}
\expandafter\ifx\csname url\endcsname\relax
  \def\url#1{\texttt{#1}}\fi
\expandafter\ifx\csname urlprefix\endcsname\relax\def\urlprefix{URL }\fi
\expandafter\ifx\csname href\endcsname\relax
  \def\href#1#2{#2} \def\path#1{#1}\fi

\bibitem{BelleIIPhysicsBook}
{E. Kou et al.}, {The Belle II Physics Book}, PTEP 2019~(12) (2019) 123C01,
  [Erratum: PTEP2020,no.2,029201(2020)].
\newblock \href {http://arxiv.org/abs/1808.10567} {\path{arXiv:1808.10567}},
  \href {http://dx.doi.org/10.1093/ptep/ptz106, 10.1093/ptep/ptaa008}
  {\path{doi:10.1093/ptep/ptz106, 10.1093/ptep/ptaa008}}.

\bibitem{BelleIIZprime}
{I. Adachi et al.}, {Search for an Invisibly Decaying ${Z}^{\ensuremath{'}}$
  Boson at Belle II in
  ${e}^{+}{e}^{\ensuremath{-}}\ensuremath{\rightarrow}{\ensuremath{\mu}}^{+}{\ensuremath{\mu}}^{\ensuremath{-}}({e}^{\ifmmode\pm\else\textpm\fi{}}{\ensuremath{\mu}}^{\ensuremath{\mp}})$
  Plus Missing Energy Final States}, {Phys. Rev. Lett.} {124} ({2020})
  {141801}.
\newblock \href {http://dx.doi.org/{10.1103/PhysRevLett.124.141801}}
  {\path{doi:{10.1103/PhysRevLett.124.141801}}}.

\bibitem{Storey}
{R. S. Storey, W. Jack and A. Ward}, {The Fluorescent Decay of {CsI}(Tl) for
  Particles of Different Ionization Density}, Proceedings of the Physical
  Society 72~(1) (1958) 1-- 8.
\newblock \href {http://dx.doi.org/10.1088/0370-1328/72/1/302}
  {\path{doi:10.1088/0370-1328/72/1/302}}.

\bibitem{Longo_2018}
{S. Longo and J. M. Roney}, Hadronic vs. electromagnetic pulse shape
  discrimination in {CsI(Tl)} for high energy physics experiments, Journal of
  Instrumentation 13~(03) (2018) P03018.
\newblock \href {http://dx.doi.org/10.1088/1748-0221/13/03/p03018}
  {\path{doi:10.1088/1748-0221/13/03/p03018}}.

\bibitem{Voss2014}
{P. Voss et al.}, {The TIGRESS Integrated Plunger ancillary systems for
  electromagnetic transition rate studies at TRIUMF}, {Nucl. Instrum. Meth.}
  {A746} ({2014}) {87 -- 97}.
\newblock \href
  {http://dx.doi.org/{https://doi.org/10.1016/j.nima.2014.02.006}}
  {\path{doi:{https://doi.org/10.1016/j.nima.2014.02.006}}}.

\bibitem{Skulski}
{W. Skulski and M. Momayezi}, {Particle identification in CsI(Tl) using digital
  pulse shape analysis}, Nucl. Instrum. Meth. A458~(3) (2001) 759 -- 771.
\newblock \href
  {http://dx.doi.org/https://doi.org/10.1016/S0168-9002(00)00938-4}
  {\path{doi:https://doi.org/10.1016/S0168-9002(00)00938-4}}.

\bibitem{AMPHORA}
{D. Drain et al.}, The particle detector array amphora, Nucl. Instrum. Meth.
  A281~(3) (1989) 528 -- 538.
\newblock \href
  {http://dx.doi.org/https://doi.org/10.1016/0168-9002(89)91487-3}
  {\path{doi:https://doi.org/10.1016/0168-9002(89)91487-3}}.

\bibitem{CHIMERA}
{M. Alderighi et al.}, Particle identification method in the csi(tl)
  scintillator used for the chimera 4$\pi$ detector, Nucl. Instrum. Meth.
  A489~(1) (2002) 257 -- 265.
\newblock \href
  {http://dx.doi.org/https://doi.org/10.1016/S0168-9002(02)00800-8}
  {\path{doi:https://doi.org/10.1016/S0168-9002(02)00800-8}}.

\bibitem{Bendel2013}
{M. Bendel et al.}, {RPID ---A new digital particle identification algorithm
  for CsI(Tl) scintillators}, {The European Physical Journal A} {49}~({6})
  ({2013}) {69}.
\newblock \href {http://dx.doi.org/{10.1140/epja/i2013-13069-8}}
  {\path{doi:{10.1140/epja/i2013-13069-8}}}.

\bibitem{Bartle}
{C.M. Bartle and R.C. Haight}, {Small inorganic scintillators as neutron
  detectors}, {Nucl. Instrum. Meth.} {A422}~({1}) ({1999}) {54 -- 58}.
\newblock \href {http://dx.doi.org/{10.1016/S0168-9002(98)01062-6}}
  {\path{doi:{10.1016/S0168-9002(98)01062-6}}}.

\bibitem{McLean2006}
{T.D. McLean et al.}, {CHELSI: Recent developments in the design and
  performance of a high-energy neutron spectrometer}, {Nucl. Instrum. Meth.}
  {A562}~({2}) ({2006}) {793 -- 796}, {Proceedings of the 7th International
  Conference on Accelerator Applications}.
\newblock \href
  {http://dx.doi.org/{https://doi.org/10.1016/j.nima.2006.02.057}}
  {\path{doi:{https://doi.org/10.1016/j.nima.2006.02.057}}}.

\bibitem{Ashida2018}
{Y. Ashida et al.}, {Separation of gamma-ray and neutron events with CsI(Tl)
  pulse shape analysis}, {Progress of Theoretical and Experimental Physics}
  {2018}~({4}), {043H01}.
\newblock \href {http://dx.doi.org/{10.1093/ptep/pty040}}
  {\path{doi:{10.1093/ptep/pty040}}}.

\bibitem{belle2002}
{A. Abashian et al.}, {The Belle Detector}, Nucl. Instrum. Meth. A479 (2002)
  117--232.
\newblock \href {http://dx.doi.org/10.1016/S0168-9002(01)02013-7}
  {\path{doi:10.1016/S0168-9002(01)02013-7}}.

\bibitem{babar2002}
{B. Aubert et al.}, {The BaBar detector}, Nucl. Instrum. Meth. A479 (2002)
  1--116.
\newblock \href {http://arxiv.org/abs/hep-ex/0105044}
  {\path{arXiv:hep-ex/0105044}}, \href
  {http://dx.doi.org/10.1016/S0168-9002(01)02012-5}
  {\path{doi:10.1016/S0168-9002(01)02012-5}}.

\bibitem{babar2013}
{B. Aubert et al.}, {The BABAR Detector: Upgrades, Operation and Performance},
  Nucl. Instrum. Meth. A729 (2013) 615--701.
\newblock \href {http://arxiv.org/abs/1305.3560} {\path{arXiv:1305.3560}},
  \href {http://dx.doi.org/10.1016/j.nima.2013.05.107}
  {\path{doi:10.1016/j.nima.2013.05.107}}.

\bibitem{besiii2009}
{M. Ablikim et al.}, {Design and Construction of the BESIII Detector}, Nucl.
  Instrum. Meth. A614 (2010) 345--399.
\newblock \href {http://arxiv.org/abs/0911.4960} {\path{arXiv:0911.4960}},
  \href {http://dx.doi.org/10.1016/j.nima.2009.12.050}
  {\path{doi:10.1016/j.nima.2009.12.050}}.

\bibitem{BelleIILumi}
{F. Abudin{\'{e}}n et al.}, {Measurement of the integrated luminosity of the
  Phase 2 data of the Belle {II} experiment}, {Chinese Physics C} {44}~({2})
  (2020) {021001}.
\newblock \href {http://dx.doi.org/{10.1088/1674-1137/44/2/021001}}
  {\path{doi:{10.1088/1674-1137/44/2/021001}}}.

\bibitem{BelleIITDR}
{T. Abe et al.}, {Belle II Technical Design Report} (2010).
\newblock \href {http://arxiv.org/abs/1011.0352} {\path{arXiv:1011.0352}}.

\bibitem{EMCalBelleII}
{V. Aulchenko et al.}, {Electromagnetic calorimeter for Belle II}, Journal of
  Physics: Conference Series 587~(1) (2015) 012045.

\bibitem{Longo_thesis}
S.~Longo, {First application of CsI(Tl) pulse shape discrimination at an $e^+$
  $e^-$ collider to improve particle identification at the Belle II
  experiment}, Ph.D. thesis, University of Victoria,
  http://hdl.handle.net/1828/11301 (10 2019).

\bibitem{GEANT4}
{S. Agostinelli et al.}, Geant4-a simulation toolkit, Nucl. Instrum. Meth.
  A506~(3) (2003) 250 -- 303.
\newblock \href
  {http://dx.doi.org/https://doi.org/10.1016/S0168-9002(03)01368-8}
  {\path{doi:https://doi.org/10.1016/S0168-9002(03)01368-8}}.

\bibitem{BirksTheoryandPractice}
{J.B. Birks}, {The theory and practice of scintillation counting}, Vol.~{27},
  {Macmillan}, {New York}, {1964}.

\bibitem{mesonInter}
{K. M. Eisenberg and D.S. Kolton}, Theory of meson interactions with nuclei,
  United States John Wiley \& Sons Inc, 1980.

\bibitem{pdg}
{C. Patrignani et al.}, {Review of Particle Physics}, Chin. Phys. C40~(10)
  (2016) 100001.
\newblock \href {http://dx.doi.org/10.1088/1674-1137/40/10/100001}
  {\path{doi:10.1088/1674-1137/40/10/100001}}.

\bibitem{CALIFA}
{B. Pietras et al.}, First testing of the califa barrel demonstrator, Nucl.
  Instrum. Meth. A814 (2016) 56 -- 65.
\newblock \href {http://dx.doi.org/https://doi.org/10.1016/j.nima.2016.01.032}
  {\path{doi:https://doi.org/10.1016/j.nima.2016.01.032}}.

\bibitem{Zeitlin2016}
{C. Zeitlin et al.}, {Calibration and Characterization of the Radiation
  Assessment Detector (RAD) on Curiosity}, {Space Science Reviews} {201}~({1})
  ({2016}) {201-- 233}.
\newblock \href {http://dx.doi.org/{10.1007/s11214-016-0303-y}}
  {\path{doi:{10.1007/s11214-016-0303-y}}}.

\bibitem{Koba2011}
{Y. Koba et al.}, {Scintillation Efficiency of Inorganic Scintillators for
  Intermediate-Energy Charged Particles}, {Prog. in Nuc. Sci. and Tech.} {1}
  ({2011}) {218--221}.

\bibitem{GwinMurray1963}
{R. Gwin and R. B. Murray}, {Scintillation Process in CsI(Tl). I. Comparison
  with Activator Saturation Model}, {Phys. Rev.} {131} ({1963}) {501--508}.
\newblock \href {http://dx.doi.org/{10.1103/PhysRev.131.501}}
  {\path{doi:{10.1103/PhysRev.131.501}}}.

\bibitem{Klempt2}
{E. Klempt et al.}, {Antinucleon-nucleon interaction at low energy:
  Annihilation dynamics}, Physics Reports 413 (2005) 197 -- 317.

\bibitem{Keck2017}
T.~Keck, {FastBDT: A Speed-Optimized Multivariate Classification Algorithm for
  the Belle II Experiment}, Comput. Softw. Big Sci. 1~(1) (2017) 2.
\newblock \href {http://dx.doi.org/10.1007/s41781-017-0002-8}
  {\path{doi:10.1007/s41781-017-0002-8}}.

\bibitem{SebastianThesis}
{S. Stengel}, {Optimization of the $\pi^0$ reconstruction selections for the
  Belle~II experiment}, Master's thesis, {Johannes Gutenberg-Universität
  Mainz} ({2019}).

\bibitem{ZERNIKE1934689}
{F. von Zernike}, {Beugungstheorie des Schneidenverfahrens und seiner
  verbesserten Form, der Phasenkontrastmethode}, {Physica} {1}~({7}) ({1934})
  {689 -- 704}.

\bibitem{PSD_CsIBaF}
{L. E. Dinca et al.}, {Alpha-gamma pulse shape discrimination in CsI:Tl, CsI:Na
  and BaF2 scintillators}, Nucl. Instrum. Meth. A486~(1) (2002) 141 -- 145.
\newblock \href {http://dx.doi.org/10.1016/S0168-9002(02)00691-5}
  {\path{doi:10.1016/S0168-9002(02)00691-5}}.

\bibitem{PSD_CsIpure}
{J. Woo et al.}, {A Pulse Shape Discrimination Method with CsI Using the Ratio
  of Areas for Identifying Neutrons and Gamma Rays}, Journal of the Korean
  Physical Society 62~(5) (2013) 839 -- 844.

\bibitem{PSD_PbWO}
{L. Bardelli et al.}, {Pulse-shape discrimination with PbWO4 crystal
  scintillators}, Nucl. Instrum. Meth. A584~(1) (2008) 129 -- 134.
\newblock \href {http://dx.doi.org/10.1016/j.nima.2007.10.021}
  {\path{doi:10.1016/j.nima.2007.10.021}}.

\end{thebibliography}

\end{document}